\documentclass[reprint,amsmath,amssymb,aip,superscriptaddress,longbibliography]{revtex4-2}
\pdfoutput=1

\usepackage{hyperref,url}
\usepackage{amsmath}
\usepackage{amsfonts}
\usepackage{braket}
\usepackage{siunitx}
\usepackage[capitalise]{cleveref}
\usepackage{placeins}
\hypersetup{breaklinks,colorlinks=true,linkcolor=blue,citecolor=blue,urlcolor=blue,filecolor=blue}
\usepackage{graphicx}
\usepackage{dcolumn}
\usepackage{tikz}
\usepackage[inline]{enumitem}
\usetikzlibrary{arrows.meta, positioning, shapes, shadows, trees}
\tikzset{
    basic/.style  = {draw, text width=5cm, drop shadow, font=\sffamily, rectangle},
    root/.style   = {basic, rounded corners=2pt, thin, align=center,
                     fill=green!60},
    level 2/.style = {basic, rounded corners=6pt, thin,align=center, fill=green!30,
                     text width=9.5em},
    level 3/.style = {basic, thin, align=left, fill=pink!60, text width=6.5em}
}
\usepackage{algorithm}
\usepackage{mhchem}
\usepackage{algpseudocode}
\usepackage{color}
\usepackage{listings,lstautogobble}
\usepackage[frozencache,cachedir=minted-cache]{minted} 
\usepackage{xcolor}
\definecolor{LightGray}{gray}{0.9}

\newminted[mycode]{python}{
    bgcolor=LightGray,
    fontsize=\footnotesize,
    breaklines=true
}

\newcommand{\ipie}{\texttt{ipie}}

\newcommand\mat\mathbf

\newcommand{\trexio}{\textsc{TrexIO}}

\newcommand{\add}[1]{{\color{black}{#1}}}

\usepackage{ulem}

\newcommand{\Google}{\affiliation{
Google Research, Venice, CA 90291, United States}}
\newcommand{\Harvard}{\affiliation{Department of Chemistry and Chemical Biology, Harvard University, Cambridge, MA, USA}}
\newcommand{\Columbia}{\affiliation{Department of Chemistry, Columbia University, New York, NY, USA}}
\newcommand{\CNRS}
{\affiliation{Laboratoire de Chimie et Physique Quantiques (UMR 5626), Universit\'e de Toulouse, CNRS, UPS, France}}
\begin{document}
\author{Tong Jiang}
\Harvard
\author{Moritz K. A. Baumgarten}
\Harvard
\author{Pierre-Fran\c{c}ois Loos}
\CNRS
\author {Ankit Mahajan}
\Columbia
\author{Anthony Scemama}
\CNRS
\author{Shu Fay Ung}
\Columbia
\author{Jinghong Zhang}
\Harvard
\author{Fionn D Malone}
\Google
\author{Joonho Lee}
\email{joonholee@g.harvard.edu}
\Harvard

\title{Improved Modularity and New Features in \texttt{ipie}: 
Towards Even Larger AFQMC Calculations
on CPUs and GPUs at Zero and Finite Temperatures
}
\begin{abstract}
\texttt{ipie} is a Python-based auxiliary-field quantum Monte Carlo (AFQMC) package that has undergone substantial improvements since its initial release [\textit{J. Chem. Theory Comput.}, 2023, 19(1): 109-121].
This paper outlines the improved modularity and new capabilities implemented in \ipie{}. 
We highlight the ease of incorporating different trial and walker types and the seamless integration of \ipie{} with external libraries.
We enable distributed Hamiltonian simulations of large systems
that otherwise would not fit on single CPU node or GPU card.
This development enabled us to compute the interaction energy of a benzene dimer with 84 electrons and 1512 orbitals with multi-GPUs. 
\add{Using \texttt{CUDA} and \texttt{cupy} for NVIDIA GPUs,}
\texttt{ipie} supports GPU-accelerated multi-slater determinant trial wavefunctions [arXiv:2406.08314] to enable efficient and highly accurate simulations of large-scale systems. This allows for near-exact ground state energies of multi-reference clusters, [Cu$_2$O$_2$]$^{2+}$ and [Fe$_2$S$_2$(SCH$_3$)$_4$]$^{2-}$.
We also describe implementations of free projection AFQMC, finite temperature AFQMC, AFQMC for electron--phonon systems, and automatic differentiation in AFQMC for calculating physical properties.
These advancements position \texttt{ipie} as a leading platform for AFQMC research in quantum chemistry, facilitating more complex and ambitious computational method development and their applications.
\end{abstract}
\maketitle

\section{Introduction}
Auxiliary-field quantum Monte Carlo (AFQMC)~\cite{zhang_constrained_1995,Zhang2003Apr} has become increasingly popular in quantum chemistry~\cite{lee_twenty_2022,motta_ab_2018}
and is furthermore being recognized as a useful approach in the context of quantum algorithms.~\cite{huggins_unbiasing_2022,wan2023matchgate,amsler2023classical,kiser2023classical,Huang2024Apr,Kiser2024Aug,Jiang2024Jul}
A well-maintained AFQMC program with flexibility and robust performance will play a pivotal role at the intersection of many disciplines, including chemistry, physics, materials science, and quantum information science.
Given the initial \texttt{PAUXY} development effort for rapid prototyping, while lacking performance,~\cite{lee2019auxiliary,lee_stochastic_2020,lee2021phaseless,lee2021constrained,lee2021spectral} the Python-based AFQMC program \texttt{ipie} was designed from scratch for high performance and ease of development. It was officially introduced as a production-level package in Ref.~\citenum{malone_ipie_2023},
\add{under the Apache License 2.0.}
\texttt{ipie} was optimized for high-performance computing architectures with both central and graphical processing
units (CPUs and GPUs). High performance was largely achieved by integrating \texttt{Numba}'s JIT compilation~\cite{lam2015proceedings} to fine-tune the computational efficiency of specific kernels and MPI parallelism for effective distributed computing.
The utility of \texttt{ipie} was showcased through the resolution of intricate quantum chemical challenges, notably the [Cu$_2$O$_2$]$^{2+}$ torture track.~\cite{cramer2006theoretical,malone_ipie_2023}
Rigorous benchmarks on CPU and GPU platforms position \texttt{ipie} competitively, displaying speed on par with--or surpassing--existing Python and C++ codes.~\cite{malone_ipie_2023,kent2020qmcpack,sharma2017semistochastic}
\add{Systematic timing benchmarks in Ref.~\citenum{malone_ipie_2023} demonstrated the cost of typical AFQMC calculations using \texttt{ipie}, also compared to other available C++ codes.
}
Since its release, \texttt{ipie} has \add{gained popularity beyond the original developer group, from academic researchers in quantum chemistry and physics to engineers in the technology sector who apply quantum simulations for material design and algorithm development.~\cite{hehn2023chelate,vysotskiy2024scalar,chen2023data,lee_twenty_2022,kiser2023classical,amsler2023classical,jiang2024,Kiser2024Aug}}

In this article, we describe the recent development and current status of 
\texttt{ipie} \add{(version v0.7.1),~\cite{ipie_github}} introducing enhanced modularity, a suite of new features, interfaces to external packages, and associated numerical examples.
Below are the key highlights:
\paragraph{\textbf{High degree of modularity and customizability.}}
The AFQMC driver has been restructured to be fully modular, allowing for a straightforward combination of features. This provides greater flexibility in adapting to a wide range of user demands.
Advancements in AFQMC algorithms often focus on developing new trial wavefunctions to control better the fermionic sign/phase problem.~\cite{huggins_unbiasing_2022,mahajan2021taming,vitali_calculating_2019,chang_auxiliary-field-based_2016,jiang2024}
The key routines in an AFQMC calculation compute intermediates using walker and trial wavefunctions, including overlap, force bias, Green's function, and energy estimators.
Without altering the internal core code of \texttt{ipie}, users and developers can customize all components making up an AFQMC simulation, including trial wavefunctions, walkers, Hamiltonians, propagators, and estimators.
The key objects are all structured using object-oriented programming (OOP) principles, facilitating straightforward customization through inheritance.
\paragraph{\textbf{Development-friendly design.}}
This improvement standardizes the component interfaces and workflow processes while offering a flexible system that adapts to various data types and user requirements. 
The abstract base classes serve as a foundational blueprint, ensuring all components adhere to a uniform structure and interact seamlessly. Complementing this, the factory methods for common workflows simplify the instantiation process, allowing users to set up standard calculations with minimal effort and reduced potential for errors. 
Integrating type-based dispatch through \texttt{Plum}~\cite{plum} brings increased precision and efficiency in method handling, ensuring every component downstream dynamically adjusts its operations based on the specific trials and walkers requested. 

Moreover, we have simplified the integration of \texttt{ipie} with external quantum chemistry packages. While the necessary integrals and orbitals for running AFQMC are most commonly obtained through an interface with \texttt{PySCF},~\cite{sun2018pyscf} our simplified file format also ensures that other packages can be easily used.
For sophisticated trial wavefunctions, such as multiple Slater determinant (MSD) trials derived from selected configuration interaction, interfaces with \texttt{PySCF},~\cite{sun2018pyscf} \texttt{Dice},~\cite{holmes2016heat,sharma2017semistochastic} and \trexio{}\cite{trexio} are available. 
An additional interface with the Fermionic Quantum Emulator (FQE)~\cite{fqe_2021,*rubin2021fermionic} has also been introduced, facilitating the conversion between \texttt{ipie}'s MSD wavefunction and quantum circuit wavefunctions, which facilitates AFQMC applications in the quantum information science (QIS) community.
\paragraph{\textbf{New improvements and features.}} \add{In the previous release paper,~\cite{malone_ipie_2023}
we introduced basic capabilities 
to perform phaseless AFQMC calculations
with both single Slater determinant and multi-Slater determinant trial wavefunctions. 
It supported CPU and GPU runs for AFQMC with single determinant trial wavefunctions, while only CPUs were supported for multi-Slater determinant calculations.
In this release, several features have been added to \texttt{ipie} to 
improve memory management, accelerate performance, handle new problems, 
enhance integration testing, among other advancements.}
\add{Notablly,
\begin{enumerate}[label=\textnormal{}]
    \item \textbf{Tackling larger systems:} To manage high storage requirement of Cholesky vectors (i.e., $\mathcal O(N^3)$,)
    \texttt{ipie} offers shared memory across MPI processes on the same node and 
    distributed-memory among CPUs and GPUs. 
    \item \textbf{Faster calculations for systems with multireference character:} \ipie{} adds support 
    for GPU-accelerated AFQMC calculation with multi-Slater determinant trials, which enables more efficient calculations 
    of the ground state of systems with multireference characters, for example, bond breaking problems~\cite{lee_twenty_2022} and 
    strongly correlated systems such as transition metal complexes.~\cite{cramer2006theoretical,malone_ipie_2023,li2017spin,hehn2023chelate}
    \item \textbf{New AFQMC developments.} While primarily developed for phaseless AFQMC (ph-AFQMC) targeting \textit{ab initio} systems, \texttt{ipie} also supports additional AFQMC methods, 
    including free-projection AFQMC,~\cite{mahajan2021taming} finite temperature AFQMC,~\cite{lee2021phaseless} AFQMC 
    for coupled electron--phonon models,~\cite{lee2021constrained} and automatic differentiation 
    within AFQMC for calculating observables that do not commute with the Hamiltonian.~\cite{mahajan2023response} 
    We also support complex-valued Hamiltonians that will be useful for performing prototypical solid-state calculations. 
    \item \textbf{Integrated testing.} \texttt{ipie} is equipped with 
    improved integration testing, boosting the package's robustness and adaptability and significantly enhancing its reliability and usability for end-users and developers. 
\end{enumerate}
}

The organization of this paper is as follows: \cref{sec:afqmc} overviews the theory of AFQMC; \cref{sec:mod} details the components, software architecture, and workflow of \texttt{ipie}, including examples to illustrate the framework's adaptability for AFQMC development;
\cref{sec:feature} introduces new features in \texttt{ipie} and provides corresponding examples;
\cref{sec:interfaces} outlines the interfaces to external packages; and
\cref{sec:conclusion} concludes with a summary and outlook.

\section{Theory of AFQMC}\label{sec:afqmc}
AFQMC is based on the following imaginary time evolution:
\begin{equation}
    \left|\Psi_0\right\rangle \propto \lim _{\tau \rightarrow \infty} \exp (-\tau \hat{H})\left|\Phi_0\right\rangle=\lim _{n\rightarrow \infty}(\exp 
 (-\Delta \tau \hat{H}))^n|\Phi_0\rangle, \label{eq:afqmc}
\end{equation}
where $\Delta\tau$ is an infinitesimal time step, $\left|\Psi_0\right\rangle$ is the ground state wavefunction, and $\left|\Phi_0\right\rangle$ is an initial state satisfying $\langle\Phi_0|\Psi_0\rangle\neq 0$.
While \texttt{ipie} supports some of the prototypical model Hamiltonians, its development has focused on the simulation of the \textit{ab initio} Hamiltonian, which in second quantization is given by
\begin{equation}
\hat{H}=\sum_{p,q=1}^N h_{p q} \hat{a}_p^{\dagger} \hat{a}_q+\frac{1}{2} \sum_{p,q,r,s=1}^N g_{psqr} \hat{a}_p^{\dagger} \hat{a}_q^{\dagger} \hat{a}_r \hat{a}_s, \label{eq:qcham}
\end{equation}
where the two-electron repulsion integral (ERI) is factorized with the Cholesky decomposition
\begin{equation}\label{eq:chol_decomp}
    g_{psqr}=(ps|qr)=\sum_{\gamma=1}^{N_\gamma} L_{ps}^\gamma L_{qr}^{\gamma}.
\end{equation}
With this factorization, we have
\begin{equation}
    \hat{H}= \hat{v}_0-\frac{1}{2} \sum_{\gamma=1}^{N_\gamma} \hat{v}_\gamma^2,
\end{equation}
where
\begin{equation}
    \hat{v}_0 = \sum_{pq} \left[h_{pq} -\frac{1}{2}\sum_r (pr|rq)\right] \hat{a}_{p}^\dagger \hat{a}_{q}
\end{equation}
\begin{equation}
\hat{v}_\gamma=\mathrm{i} \sum_{p q} L_{p q}^\gamma \hat{a}_{p }^{\dagger} \hat{a}_{q}.   \label{eq:chol}
\end{equation}
The short-time propagator with Trotter decomposition is written as
\begin{equation}
    \mathrm{e}^{-\Delta \tau \hat{H}}=\mathrm{e}^{-\frac{\Delta\tau}{2} \hat{v}_0} \mathrm{e}^{\frac{\Delta\tau}{2} \sum\hat{v}_{\gamma}^2} \mathrm{e}^{- \frac{\Delta\tau}{2} \hat{v}_0}+\mathcal{O}\left(\Delta \tau^3\right). \label{eq:trotter}
\end{equation}
Upon applying the Hubbard--Stratonovich transformation,\cite{hubbard1959calculation,Stratonovich} our effective propagator contains only one-body operators,
\begin{equation}
    \mathrm{e}^{-\Delta \tau \hat{H}} = \int \mathrm{d} \mathbf{x} \ p(\mathbf{x}) \hat{B}(\mathbf{x},\Delta\tau)+\mathcal{O}\left(\Delta \tau^2\right),
\end{equation}
where $p(\mathbf x)$ is the standard Gaussian distribution, $\textbf{x}=(x_1, x_2, \cdots, x_{N_{\gamma}})$ are the auxiliary fields, and the one-body propagator $\hat{B}$ is
\begin{equation}
    \hat{B}(\mathbf{x},\Delta\tau) = \mathrm{e}^{-\frac{\Delta\tau}{2} \hat{v}_0} \mathrm{e}^{-\sqrt{\Delta \tau} \mathbf{x}\cdot \hat{\mathbf{v}}}\mathrm{e}^{-\frac{\Delta\tau}{2} \hat{v}_0}.\label{eq:Bprop}
\end{equation}
where $\hat{\mathbf{v}}=(\hat{v}_1,\hat{v}_2,\cdots,\hat{v}_{N_\gamma})$.

\begin{figure}
    \centering
    \includegraphics[width=0.5\textwidth]{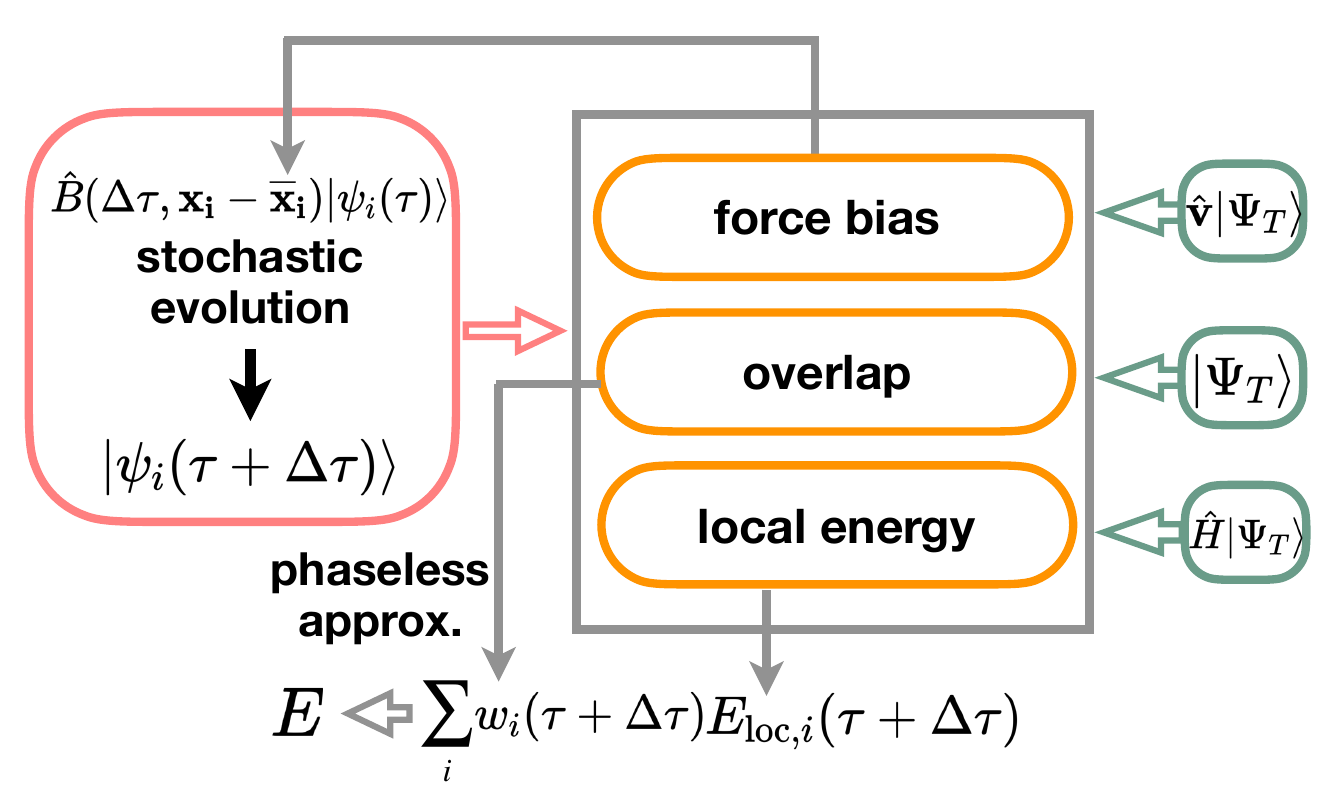}
    \caption{Overview of phaseless AFQMC in terms of computational hotspots.}
    \label{fig:afqmc}
\end{figure}

In AFQMC, each walker samples auxiliary fields and represents a statistical sample of the global wavefunction at imaginary time $\tau$, written as 
\begin{equation}
|\Psi(\tau)\rangle=\sum_i^{N_\textrm{w}} w_i(\tau) \frac{\left|\psi_i(\tau)\right\rangle}{\left\langle\Psi_\textrm{T}|\psi_i(\tau)\right\rangle},
\end{equation}
where $|\psi_i(\tau)\rangle$ is the wavefunction of the $i$-th walker at time $\tau$ and $|\Psi_\text{T}\rangle$ is the trial wavefunction used for importance sampling.
The energy estimator is then
\begin{equation}
    E(\tau) = \sum_i^{N_\textrm{w}} w_i(\tau) E_{\textrm{loc},i}(\tau)= \sum_i^{N_\textrm{w}} w_i(\tau)\frac{\langle \Psi_\textrm{T}|\hat{H}|\psi_i(\tau)\rangle}{\left\langle\Psi_\textrm{T} | \psi_i(\tau)\right\rangle}\label{eq:Eloc}.
\end{equation}
The walker state $\left|\psi_i(\tau)\right\rangle$ is updated by applying the discrete-time propagator, and the weight $w_i(\tau)$ is updated following the phaseless approximation:~\cite{Zhang2003Apr}
\begin{gather}
\left|\psi_i(\tau+\Delta \tau)\right\rangle  =\hat{B}\left(\mathbf{x}_i-\overline{\mathbf{x}}_i, \Delta \tau \right)\left|\psi_i(\tau)\right\rangle \label{eq: propagate}
\\
w_i(\tau+\Delta \tau)  =I_{\mathrm{ph}}\left(\mathbf{x}_i, \overline{\mathbf{x}}_i, \tau, \Delta \tau\right) \times w_i(\tau).  \label{eq:weight_update}
\end{gather}
We dynamically shift the distribution of auxiliary fields using the force bias $\overline{\mathbf{x}}_i$ defined by
\begin{equation}
    \overline{\mathbf{x}}_i(\Delta \tau, \tau)=-\sqrt{\Delta \tau} \frac{\left\langle\Psi_{\textrm{T}}\left|\hat{\mathbf{v}}-\langle\hat{\mathbf{v}}\rangle_\text{T}\right| \psi_i(\tau)\right\rangle}{\left\langle\Psi_{\textrm{T}}|\psi_i(\tau)\right\rangle},\label{eq:fb}
\end{equation}
where $\langle\hat{\mathbf{v}}\rangle_\text{T}$ is usually called the mean-field shift in the AFQMC literature.~\cite{motta_ab_2018} Furthermore, the phaseless importance function~\cite{zhang_constrained_1995,Zhang2003Apr} used in the weight update \eqref{eq:weight_update} is given by
\begin{equation}
    I_{\mathrm{ph}}\left(\mathbf{x}_i, \overline{\mathbf{x}}_i, \tau, \Delta \tau\right)=\left|I\left(\mathbf{x}_i, \overline{\mathbf{x}}_i, \tau, \Delta \tau\right)\right| \times \max \left[0, \cos \left(\theta_i(\tau)\right)\right],
\end{equation}
which remains real and positive throughout the propagation.
The so-called hybrid importance function is given by
\begin{equation}
    I\left(\mathbf{x}_i, \overline{\mathbf{x}}_i, \tau, \Delta \tau\right)=S_i(\tau, \Delta \tau) \mathrm{e}^{\mathbf{x}_i \cdot \overline{\mathbf{x}}_i-\overline{\mathbf{x}}_i \cdot \overline{\mathbf{x}}_i / 2},
\end{equation}
and the overlap ratio of the $i$-th walker is
\begin{equation}
    S_i(\tau, \Delta \tau)=\frac{\left\langle\Psi_{\textrm{T}}\left|\hat{B}\left(\Delta \tau, \mathbf{x}_i-\overline{\mathbf{x}}_i\right)\right| \psi_i(\tau)\right\rangle}{\left\langle\Psi_{\textrm{T}}|\psi_i(\tau)\right\rangle}. \label{eq:ovlp_ratio}
\end{equation}
We define the phase of the overlap as
\begin{equation}
    \theta_i(\tau)=\arg \left[S_i(\tau, \Delta \tau)\right].
\end{equation}
The computation of local energies in \cref{eq:Eloc}, propagation of wavefunctions in \cref{eq: propagate}, and evaluation of force biases in \cref{eq:fb} are the primary computational hotspots in AFQMC calculations. We optimize these routines by employing optimized algorithms and adopting high-performance computing techniques. We provide an overview of the phaseless AFQMC algorithm described above in Fig.~\ref{fig:afqmc}.

\section{Software architecture and design principles}\label{sec:mod}
\subsection{AFQMC driver}
\begin{figure*}
    \centering
    \includegraphics[width=0.9\textwidth]{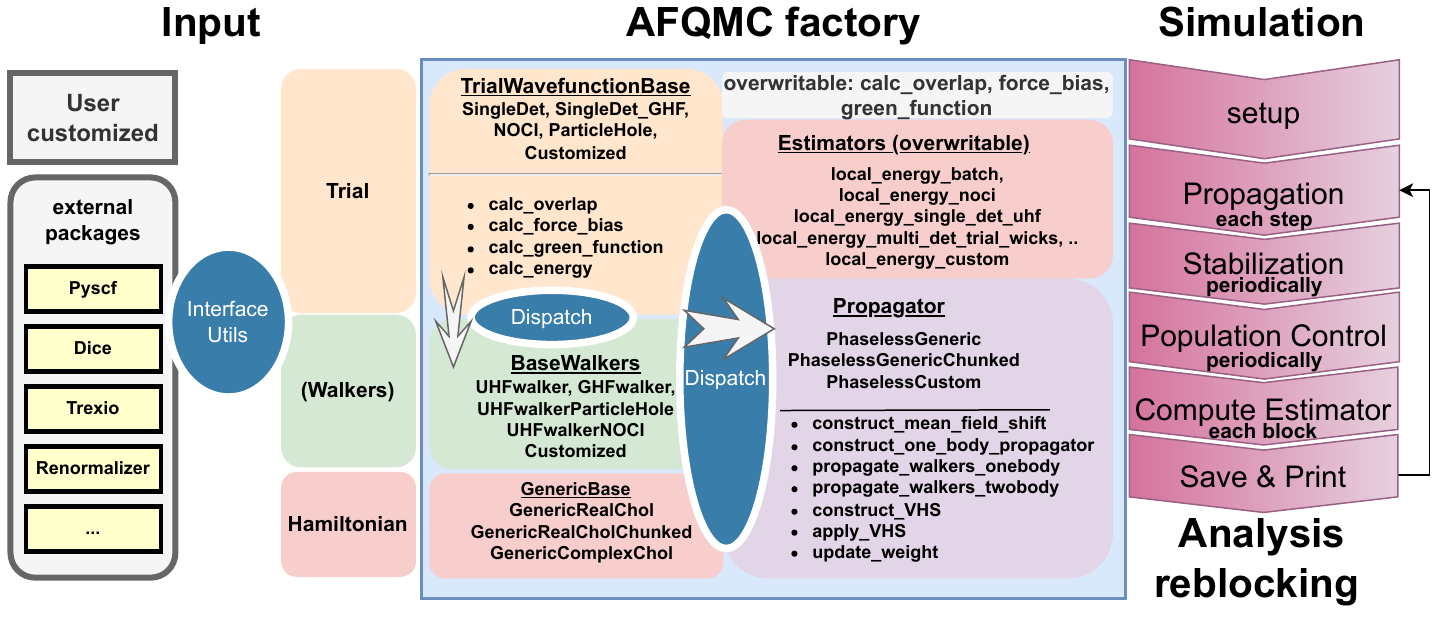}
    \caption{The workflow of \texttt{ipie}.}
    \label{fig:ipie_wflow}
\end{figure*}

We overview the software architecture of \texttt{ipie} with emphasis on improved modularity, as shown in Fig.~\ref{fig:ipie_wflow}.
\texttt{ipie} generates an AFQMC simulation by assembling the generic AFQMC driver from problem-specific components and then running it to execute the simulation.
The driver requires several inputs to define the QMC simulation, including the trial wavefunction, walkers, and the Hamiltonian, as well as more generic parameters such as the number of blocks and time step. 
The following Python Code Snippet shows how the driver can be instantiated:
\begin{listing}[H]
\begin{minted}[tabsize=2,breaklines,fontsize=\footnotesize,bgcolor=LightGray]{python}
class AFQMC(object):
    def __init__(self, system, hamiltonian, trial, 
        walkers,propagator, mpi_handler, params:
        QMCParams):
        ......
\end{minted}
\caption{The AFQMC class within \texttt{ipie}.}
\label{lst:afqmc-class}
\end{listing}
The code takes several objects needed for a typical AFQMC driver:
\begin{enumerate}[label=(\arabic*)]
    \item \texttt{system} contains information that defines the problem, including the number of spin-up and spin-down electrons.
    \item \texttt{hamiltonian} stores molecular integrals as detailed in Sec.~\ref{sec:Hamiltonian}.
    \item \texttt{trial} represents a trial wavefunction object, for which \texttt{ipie} provides many options as detailed in Sec.~\ref{sec:trial}.
    \item \texttt{walkers} manages all information about walkers during the imaginary time propagation. This includes the initial walker wavefunctions and the number of walkers, among other details. \texttt{ipie} allows walkers to be explicitly passed into the AFQMC driver or dynamically dispatched with default settings, as detailed in Sec.~\ref{sec:plum}.
    \item \texttt{propagator} handles the update of walker weights and wavefunctions during the imaginary time evolution. \texttt{ipie}'s AFQMC driver can take the propagator class as an input or dispatch it internally with default settings.
    \item \texttt{mpi\_handler} manages information related to the Message Passing Interface (MPI) for parallel computing, including rank details, chunking, and groups (i.e., a group is a collection of MPI processes). The parameter \texttt{shared\_comm} in \texttt{mpi\_handler} refers to the MPI rank within a group where the chunking integrals are distributed. The specifics of these are elaborated in Section~\ref{sec:chunking}.
    \item \texttt{params} encompasses other fundamental parameters for the QMC simulation, such as the time step.
\end{enumerate}

While users can directly provide all of the aforementioned inputs to construct the AFQMC driver object, especially for development purposes, a factory method is provided to simplify the process greatly. This is especially useful for standard AFQMC calculations using single or multiple Slater determinant trials. This method significantly reduces the number of inputs required to construct the AFQMC object, as demonstrated in Code Snippet~\ref{lst:afqmc-driver}:
\begin{listing}[H]
    \begin{mycode}
    def build(
        num_elec: Tuple[int, int],
        hamiltonian,
        trial_wavefunction
    ) -> "AFQMC":
        ......
    \end{mycode}
    \caption{Factory method for AFQMC driver.}
    \label{lst:afqmc-driver}
\end{listing}

As illustrated in the Code Snippet~\ref{lst:afqmc-driver} and Fig.~\ref{fig:ipie_wflow}, the construction of the AFQMC driver ultimately relies on two key inputs: the Hamiltonian and the trial wavefunction.
The Hamiltonian object is assembled with one-electron integrals and the Cholesky decomposition (or density fitting) of two-electron integrals, detailed in Sec.~\ref{sec:Hamiltonian}.
As for trial wavefunctions, \texttt{ipie} offers various options that utilize interfaces with external packages such as \texttt{PySCF},~\cite{sun2018pyscf} \texttt{Dice},~\cite{sharma2017semistochastic} \trexio{}~\cite{trexio} to facilitate standardized workflows. \ipie{} can also accommodate user-customized trials, which offers flexibility in future developments. 
Benefited from the flexibility of \texttt{ipie},
we are also able to incorporate more complicated trial wavefunctions, such as density matrix renormalization group (DMRG), and in Sec.~\ref{sec:trial} we show the AFQMC calculation with matrix product states (MPS) trials by incorporating \texttt{ipie} with the DMRG package \texttt{Renormalizer}.~\cite{renormalizer,ren2022time}

The \texttt{build} method in the AFQMC class is designed to streamline the setup of an AFQMC calculation by requiring only three inputs: the number of electrons, the Hamiltonian, and the trial wavefunction.
Its main purpose is automating the construction of key components such as walkers, propagators, and estimators, \textit{etc.}. Users need only specify the \texttt{hamiltonian} and trial inputs \texttt{trial\_wavefunction}, with the \texttt{build} method handling downstream instantiation of the rest. 

Once the driver is built, wavefunction propagation, stabilization, population control, and estimator calculation are performed with the provided QMC parameters. Simulation outputs in both text file and \texttt{hdf5} file formats are saved and updated after each block, which enables real-time reblocking analysis~\cite{flyvbjerg1989error} via built-in tools.


\subsection{Hamiltonian}\label{sec:Hamiltonian}
The construction of a Hamiltonian object can be achieved in several ways. 
One can construct the object directly by inputting the one-electron integrals and the Cholesky decomposition (Eq.~\eqref{eq:chol_decomp}) of either 8-fold (for real symmetric integrals) or 4-fold (for complex hermitian integrals) symmetric two-body integrals:
\begin{listing}[H]
    \begin{mycode}
from ipie.hamiltonians.generic import Generic as HamGeneric
ham = HamGeneric(h1e, chol, ecore)      
    \end{mycode}
\caption{Constructing the Hamiltonian object with provided electron integrals}
\label{lst:ham_generic}
\end{listing}
\noindent where \texttt{h1e}, \texttt{chol} and \texttt{ecore} are the one-electron integrals, the Cholesky decomposition (\cref{eq:chol}) of the two-electron integrals within the desired orbitals in \cref{eq:qcham}, and the nuclear repulsion energy, respectively.

\texttt{ipie} also offers built-in functions for assembling the Hamiltonian object from \texttt{PySCF} calculations. One can provide: (i) a \texttt{PySCF} check file; (ii) a \texttt{PySCF} \texttt{mol} object together with the molecular orbital (MO) coefficient matrix \texttt{mo\_coeff}; or (iii) a \texttt{mol} object, \texttt{mo\_coeff}, and a basis transformation matrix \texttt{X} that transforms MOs to orthogonal atomic orbitals (OAO), natural orbitals and so on, depending on the single particle basis used in the AFQMC calculation:
\begin{listing}[H]
        \begin{mycode}
from ipie.utils.from_pyscf import generate_hamiltonian, generate_hamiltonian_from_chk

# OAO basis
ham = generate_hamiltonian_from_chk(
    'scf.chk', use_mcscf=False, 
    chol_cut=1e-5, num_frozen_core=0, ortho_ao=True
)

# MO basis
ham = generate_hamiltonian(mol, mo_coeff, h1e, mo_coeff, 
    chol_cut=1e-5, num_frozen_core=0)
# basis with transform matrix X
ham = generate_hamiltonian(mol, mo_coeff, h1e, X, 
    chol_cut=1e-5)
    \end{mycode}
\caption{Constructing the Hamiltonian object from \texttt{PySCF}.}
\label{lst:from_pyscf}
\end{listing}
\noindent The argument \texttt{chol\_cut} specifies the cutoff for the Cholesky decomposition; \texttt{use\_mcscf} specifies whether to use the multi-configurational self-consistent field (MCSCF) MO coefficients; and \texttt{num\_frozen\_core} specifies the number of frozen cores for the subsequent AFQMC calculations.

Given that the size of Cholesky vectors scales as $\mathcal{O}(N^3)$, with $N$ as the number of orbitals, large systems with large basis sets can generate electron integrals of considerable size. For such systems, one cannot store copies of integrals for each MPI process and must resort to other strategies.
Utilizing shared memory across different MPI processes, \ipie{} offers an approach that allows processes on the same node to access Cholesky vectors in the same memory block. The following code snippet can be adopted to ensure the use of shared memory Hamiltonian objects:
\begin{listing}[H]
    \begin{mycode}
from ipie.utils.mpi import get_shared_comm
from ipie.hamiltonians.utils import get_hamiltonian
from mpi4py import MPI
shared_comm = get_shared_comm(MPI.COMM_WORLD)
ham = get_hamiltonian("ham.h5",comm=shared_comm,
    pack_chol=True)    
    \end{mycode}
\caption{Constructing the Hamiltonian object with shared memory.}
\label{lst:shmem}
\end{listing}
\noindent This reads the integrals from the \texttt{ham.h5} file and generates the Hamiltonian object with shared integrals across all MPI processes.
\texttt{pack\_chol} refers to another strategy used by \ipie{} to reduce memory usage by a factor of two and speed up the propagation. Namely, the permutation symmetry of Cholesky vectors is exploited to utilize only the upper-triangular part of the Cholesky vectors.
Propagation using the packed matrix contracts only the symmetric part in Eq.~\eqref{eq: propagate} and is thus more efficient.

However, the shared memory strategy often creates a memory access overhead and reduces the MPI parallel efficiency when many cores are used. 
We also enable a chunking (or distributed-memory) strategy.
The Cholesky vectors are divided into segments, or `chunked,' and distributed across different MPI processes to optimize memory usage and computational efficiency. 
The Cholesky vectors are divided into several chunks, each of which will be allocated to separate MPI processes. 
The specifics of this chunking process and the associated MPI communication strategy will be thoroughly explained in Sec.~\ref{sec:chunking}.
Given enough nodes, our approach completely removes the memory bottleneck in AFQMC and can work on both CPU and GPU nodes. We especially recommend our distributed-memory implementation for GPUs, where memory capacity is much more limited than that of CPUs. 

\subsection{Trial wavefunctions}\label{sec:trial}
\begin{figure*}
\centering
\begin{tikzpicture}[
    level 1/.style={sibling distance=40mm},
    edge from parent/.style={->,draw},
    >=latex]

\node[root] {Trial (WavefunctionBase)}
  child {node[level 2] (c1) {calc\_overlap()}}
  child {node[level 2] (c2) {calc\_force\_bias()}}
  child {node[level 2] (c3) {calc\_greens\_function()$*$}}
  child {node[level 2] (c4) {half\_rotate()$*$}};
 
\begin{scope}[every node/.style={level 3}]
\node [below of = c1, xshift=15pt] (c11) {$\langle\Psi_\textrm{T}|\psi_i\rangle$};
\node [below of = c2, xshift=15pt] (c21) {Eq.~\eqref{eq:fb}};
\node [below of = c3, xshift=15pt] (c31) {$\frac{\langle\Psi_\textrm{T}\left|\hat{a}_{p}^{\dagger} \hat{a}_{q}\right| \psi_i\rangle}{\langle\Psi_\textrm{T} \mid \psi_i\rangle}$};
\node [below of = c4, xshift=15pt] (c41) {see Ref.~\citenum{lee_stochastic_2020}};
\end{scope}

\foreach \value in {1,2,3}
  \draw[->] (c1.195) |- (c11.west);
\foreach \value in {1,2,3}
  \draw[->] (c2.195) |- (c21.west);
\foreach \value in {1,2,3}
  \draw[->] (c3.195) |- (c31.west);
\foreach \value in {1,2,3}
  \draw[->] (c4.195) |- (c41.west);
\end{tikzpicture}
\caption{The functions to be defined for a trial object. $*$: The function is optional depending on the trial type, used in certain trials to computations like overlap, force bias, or local energy.}\label{fig:trial_func}
\end{figure*}
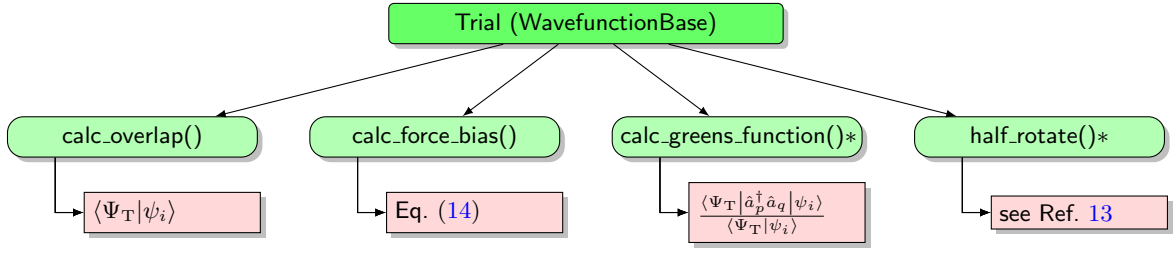

Any new trial wavefunction may inherit from the \texttt{TrialWavefunctionBase} class, which contains methods to calculate essential quantities including the overlap (Eq.~\eqref{eq:ovlp_ratio}), force bias (Eq.~\eqref{eq:fb}), \textit{etc.}, as shown in Fig.~\ref{fig:trial_func}. Some trials, such as single/multiple Slater determinants, also require computation of Green's function and half rotation of electron repulsion integrals.~\cite{lee_stochastic_2020}
There is flexibility to either inherit these functions directly or overwrite them with customized implementations, ensuring both consistency in fundamental operations and adaptability for specialized needs.

Single Slater determinant (SD) trials can be generated with
\begin{listing}[H]
    \begin{mycode}
from ipie.trial_wavefunction.single_det import SingleDet
trial = SingleDet(np.hstack([orbs_a, orbs_b]), nelec, 
    num_basis)
    \end{mycode}
\caption{The single determinant trial object.}
\label{lst:sd_trial}
\end{listing}
\noindent where \texttt{orbs\_a} and \texttt{orbs\_b} are the spin-up and spin-down coefficient matrices for the SD trials. 
Similarly, MSD trials can be generated with:
\begin{listing}[H]
    \begin{mycode}
from ipie.trial_wavefunction.particle_hole import  ParticleHole
trial = ParticleHole(wfn, nelec, num_basis)        
    \end{mycode}
\caption{The multiple Slater determinant trial object.}
\label{lst:msd_trial}
\end{listing}
\noindent with \texttt{wfn} containing the occupation indices in the basis used in the configuration interaction (CI) expansion of the MSD and their coefficients.

A new customized trial can be defined straightforwardly in \texttt{ipie}. For instance,
we consider a ``noisy'' SD trial that can be inherited from our existing SD trial class.
The noisy trial adds Gaussian noise to the overlap evaluation and keeps everything else the same as for regular SD trials.
The following code snippet achieves this:
\begin{listing}[H]
\begin{mycode}
class NoisySingleDet(SingleDet):
    def __init__(self, wavefunction, num_elec, 
        num_basis, noise_level=1e-12):
        super().__init__(wavefunction, num_elec, 
            num_basis)
        self._noise_level = noise_level
    def calc_overlap(self, walkers) -> np.ndarray:
        ovlp = super().calc_overlap(walkers)
        noise = np.random.normal(
            scale=self._noise_level, 
            size=ovlp.size
        )
        return ovlp * (1 + noise)
\end{mycode}
\caption{Example of constructing a custom noisy trial class.}
\label{lst:noise_trial}
\end{listing}

\begin{figure}
    \centering
    \includegraphics[width=0.45\textwidth]{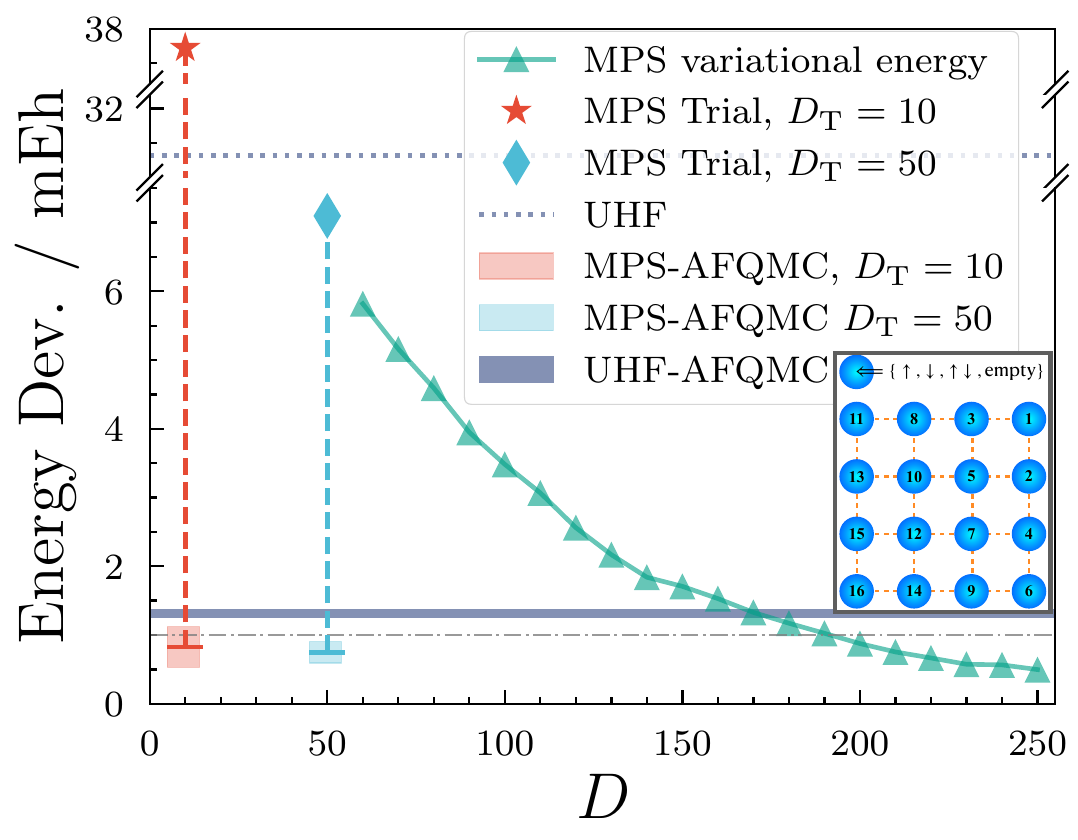}
    \caption{\textbf{Application of MPS-AFQMC to two-dimensional hydrogen lattice with $r=4.2a_0$.} Energy deviation from exact results using MPS-AFQMC, UHF-AFQMC, and DMRG with different bond dimensions.~\cite{jiang2024}}
    \label{fig:h4x4}
\end{figure}

Recently, we have also implemented a more sophisticated extension using a matrix product state (MPS) trial wavefunction.~\cite{jiang2024}
MPS is the variational ansatz used in the density matrix renormalization group (DMRG) algorithm, and we obtain the MPS solution from external DMRG packages such as \texttt{Renormalizer}.~\cite{renormalizer,*ren2022time}
The theory explaining the structure of the MPS trial and associated functions is detailed in Ref.~\citenum{jiang2024}.
A representative calculation is shown in \cref{fig:h4x4}.

\subsection{Dispatchers of walkers, propagators and estimators}\label{sec:plum}
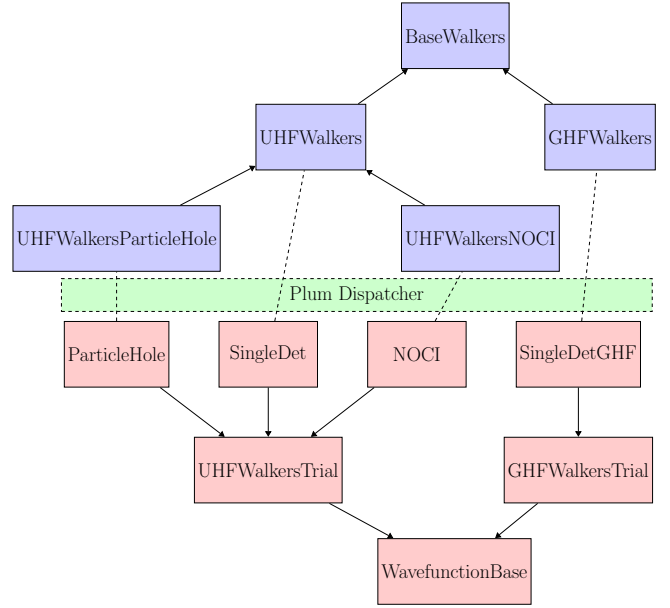
\begin{figure}
    \centering
    \resizebox{0.48\textwidth}{!}{
\begin{tikzpicture}[
    node distance=1.5cm,
    class/.style={rectangle, draw=black, thick, fill=blue!20, minimum width=3cm, minimum height=2cm, text centered,font=\fontsize{16pt}{15pt}\selectfont},
    dispatch/.style={rectangle, draw=black, thick, fill=red!20, minimum width=3cm, minimum height=2cm, text centered,font=\fontsize{16pt}{15pt}\selectfont},
    inherit/.style={->, >=Triangle, thick},
    edge/.style = {rectangle, draw=black, dashed, fill=green!20, minimum width=18cm, minimum height=1cm, text centered,font=\fontsize{16pt}{15pt}\selectfont}
]
\node[edge] at (-3, -7.9) {Plum Dispatcher};

\node (BaseWalkers) [class] {BaseWalkers};
\node (GHFWalkers) [class, below right=of BaseWalkers] {GHFWalkers};
\node (UHFWalkers) [class, below left=of BaseWalkers] {UHFWalkers};
\node (UHFWalkersParticleHole) [class, below left=of UHFWalkers] {UHFWalkersParticleHole};
\node (UHFWalkersNOCI) [class, below right=of UHFWalkers] {UHFWalkersNOCI};

\draw[inherit] (GHFWalkers) -- (BaseWalkers);
\draw[inherit] (UHFWalkers) -- (BaseWalkers);
\draw[inherit] (UHFWalkersParticleHole) -- (UHFWalkers);
\draw[inherit] (UHFWalkersNOCI) -- (UHFWalkers);

\node (ParticleHole) [dispatch, below=of UHFWalkersParticleHole] {ParticleHole};
\node (SingleDet) [dispatch, right=of ParticleHole] {SingleDet};
\node (NOCI) [dispatch, right=of SingleDet] {NOCI};
\node (SingleDetGHF) [dispatch, right=of NOCI] {SingleDetGHF};

\node (GHFWalkersTrial) [dispatch, below=of SingleDetGHF] {GHFWalkersTrial};
\node (UHFWalkersTrial) [dispatch, below=of SingleDet] {UHFWalkersTrial};
\node (WavefunctionBase) [dispatch, below right=of UHFWalkersTrial] {WavefunctionBase};

\draw[inherit] (SingleDet) -- (UHFWalkersTrial);
\draw[inherit] (SingleDetGHF) -- (GHFWalkersTrial);
\draw[inherit] (ParticleHole) -- (UHFWalkersTrial);
\draw[inherit] (NOCI) -- (UHFWalkersTrial);
\draw[inherit] (UHFWalkersTrial) -- (WavefunctionBase);
\draw[inherit] (GHFWalkersTrial) -- (WavefunctionBase);

\draw[dashed] (ParticleHole) -- (UHFWalkersParticleHole);
\draw[dashed] (SingleDet) -- (UHFWalkers);
\draw[dashed] (NOCI) -- (UHFWalkersNOCI);
\draw[dashed] (SingleDetGHF) -- (GHFWalkers);

\end{tikzpicture}
}
    \caption{Dispatcher for different walkers depending on the trial. Lines with arrows indicate class inheritance, while dashed lines signify the use of the \texttt{Plum} dispatcher for dispatching.}
    \label{fig:walker_dispatcher}
\end{figure}

As the program grows, additional complexity arises from the need to dispatch different features for arbitrary combinations of smaller building blocks. 
While there are numerous ways to achieve this, \ipie{} previously relied on conditional \texttt{if} statements at the computational subroutines, which--despite its simplicity--will result in many intractable conditions as the number of features increases.
The latest release of \texttt{ipie} thus utilizes the multiple dispatch library \texttt{Plum}~\cite{plum} to
dispatch features for different combinations of objects in a compact and flexible manner. 
This can be especially useful in AFQMC, where the behavior of many functions varies based on the type of trial, walkers, and Hamiltonians.
In Fig.~\ref{fig:walker_dispatcher}, we illustrate the dispatcher for walkers depending on the type of trial.
As for \texttt{ipie}'s native trial wavefunctions, including both single and multiple (orthogonal and non-orthogonal) Slater determinants, once the trial and Hamiltonian objects are in place, the AFQMC driver can be formed. 
In instances where the walkers object is not explicitly provided, the AFQMC driver constructs it inside the \texttt{build} function based on the type of the trial wavefunction via \texttt{Plum}.
The appropriate energy estimators and propagators are also built according to the object type of the trial, walkers, and Hamiltonian, completing the necessary initialization for AFQMC calculations.

\subsection{Estimators}
\texttt{ipie} supports well-optimized energy estimators for wavefunctions, including SD,~\cite{lee_stochastic_2020} MSD,~\cite{mahajan2022selected} and non-orthogonal configuration interaction (NOCI) trials.
It supports real and complex Hamiltonians on both CPUs and GPUs.
As mentioned in Section.~\ref{sec:plum}, dispatchers can automatically determine a specific implementation of the estimator required for given user inputs.
Users can also compose customized estimators for new types of trial wavefunctions. 
Considering the example in Code Snippet~\ref{lst:noise_trial}, we may define the corresponding energy estimator as
\begin{listing}[H]
\begin{mycode}
class NoisyEnergyEstimator(EnergyEstimator):
    def __init__(
        self, system, ham, trial):
        super().__init__(system=system, ham=ham, 
            trial=trial)
    def compute_estimator(self, system, walkers, 
        hamiltonian, trial, istep=1):
        trial.calc_greens_function(walkers)
        energy = local_energy_batch(system, hamiltonian, 
            walkers, trial)
        self._data["ENumer"] = xp.sum(walkers.weight * 
            energy[:, 0].real)
        self._data["EDenom"] = xp.sum(walkers.weight)
        self._data["E1Body"] = xp.sum(walkers.weight * 
            energy[:, 1].real)
        self._data["E2Body"] = xp.sum(walkers.weight * 
            energy[:, 2].real)
        return self.data
\end{mycode}
\caption{Customized noisy estimator.}
\label{{lst:noisy_estimator}}
\end{listing}
\noindent where \texttt{xp} serves as an abstract linear algebra backend, facilitating seamless integration of either NumPy for pure CPU-based computations or CuPy for GPU-accelerated tasks. In this estimator, we still evaluate the local energy as we do for any SD trial.
This customized estimator can be passed to the AFQMC driver simply with
\begin{listing}[H]
\begin{mycode}
add_est = {"energy": 
    NoisyEnergyEstimator(Generic(mol.nelec), ham, trial)
    }
afqmc.run(additional_estimators=add_est)
\end{mycode}
\caption{Adding the customized noisy estimator to the AFQMC driver.}
\label{lst:add_estimator}
\end{listing}
\noindent For more sophisticated cases such as MPS trials, in addition to the usual energy estimator, we can furthermore define other estimators such as one to measure the bond dimension of walkers when converted into an MPS.~\cite{jiang2024}
\section{New Features}\label{sec:feature}
Many new features have been added to \ipie{} since our first release paper;~\cite{malone_ipie_2023} we highlight some of the more important ones in this section.

\subsection{Distributed Hamiltonian for limited memory}\label{sec:chunking}
\begin{figure}
    \centering
    \includegraphics[width=0.48\textwidth]{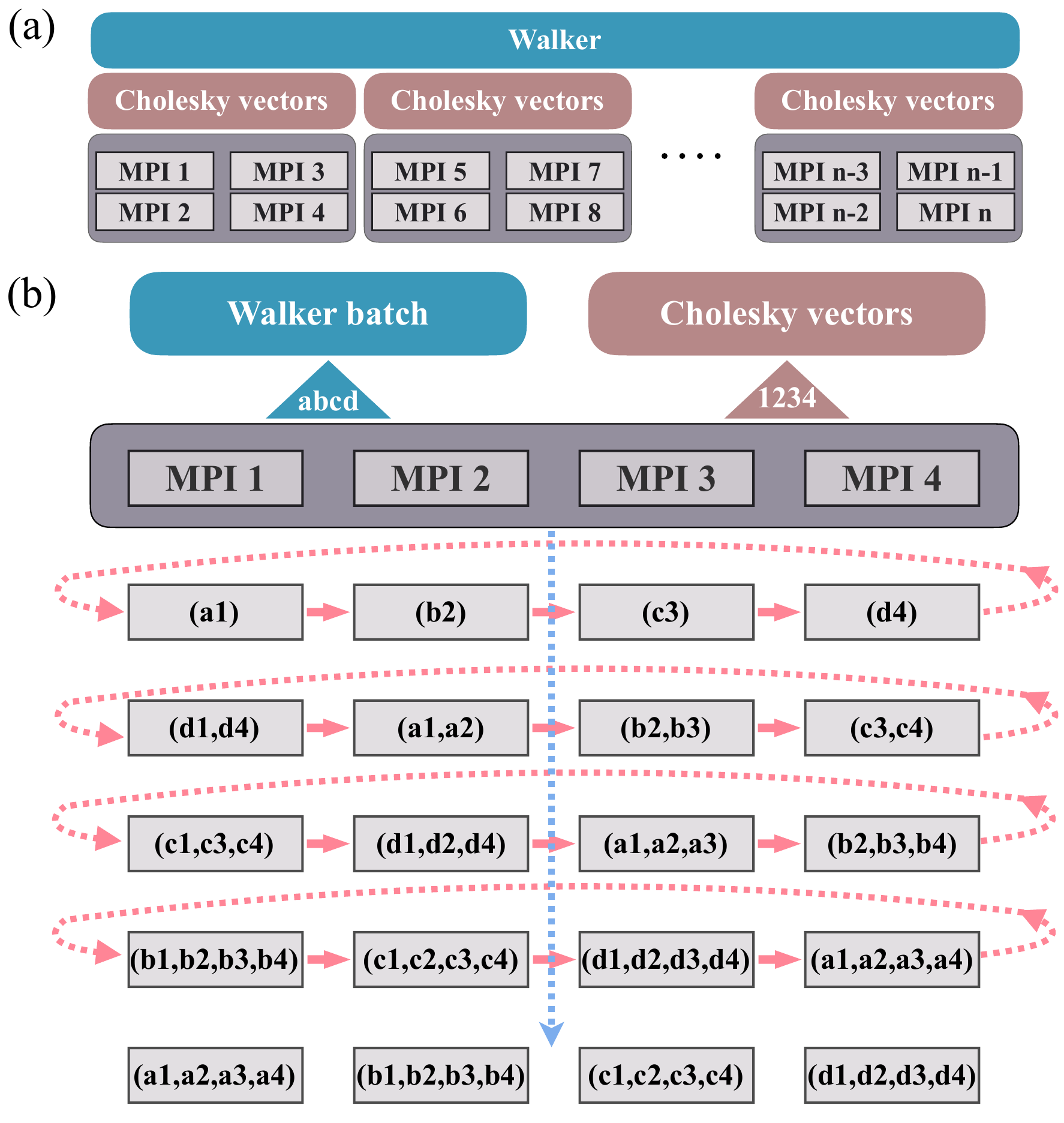}
    \caption{(a) Distribution of walkers and the chunked Hamiltonian over multiple groups. (b) MPI communication between MPI processes within a group with the distributed Hamiltonian and walker batches.}
    \label{fig:chunking_ring}
\end{figure}

When using GPUs, we routinely encounter cases where the Cholesky vectors cannot entirely fit into a single GPU card's memory.
Or one could imagine a large system where a single CPU node cannot fit the Cholesky vectors in memory anymore. 
In such circumstances, the Cholesky vectors are distributed over several GPU cards or CPU nodes that constitute a group. Each group stores the full Cholesky tensor as shown in Fig.~\ref{fig:chunking_ring}(a).
Whenever the tensor is required by the computational subroutines, for example, in computing the force bias or local energy, the walkers in each member of the group are communicated to other members in the same group, as illustrated in Fig.~\ref{fig:chunking_ring}(b).
We employ a cyclic data passing scheme of walkers among MPI processes to apply all existing Cholesky chunks to given walkers. With this strategy, one can accumulate locally contracted quantities before moving to the next QMC time step.

We performed timing benchmarks as an application by computing the parallel-displaced benzene dimer, a Van der Waals complex taken from the S22 dataset.~\cite{jurevcka2006benchmark}
Disk usage for the benzene dimer calculations varies with the choice of basis set: 11GB for the aug-cc-pVTZ basis and 67GB for the aug-cc-pVQZ basis with a Cholesky decomposition threshold of $10^{-3}$. 
The threshold is set based on the Hartree-Fock energy error around typical density fitting errors~\cite{weigend2009approximated} ($< 50\mu E_\textrm{h}$ per atom here).
We used our shared memory framework for CPU calculations and distributed Cholesky vectors and half-rotated integrals~\cite{lee_stochastic_2020} across multiple cards were used for GPU computations.

\begin{figure}
    \centering
    \includegraphics[width=0.5\textwidth]{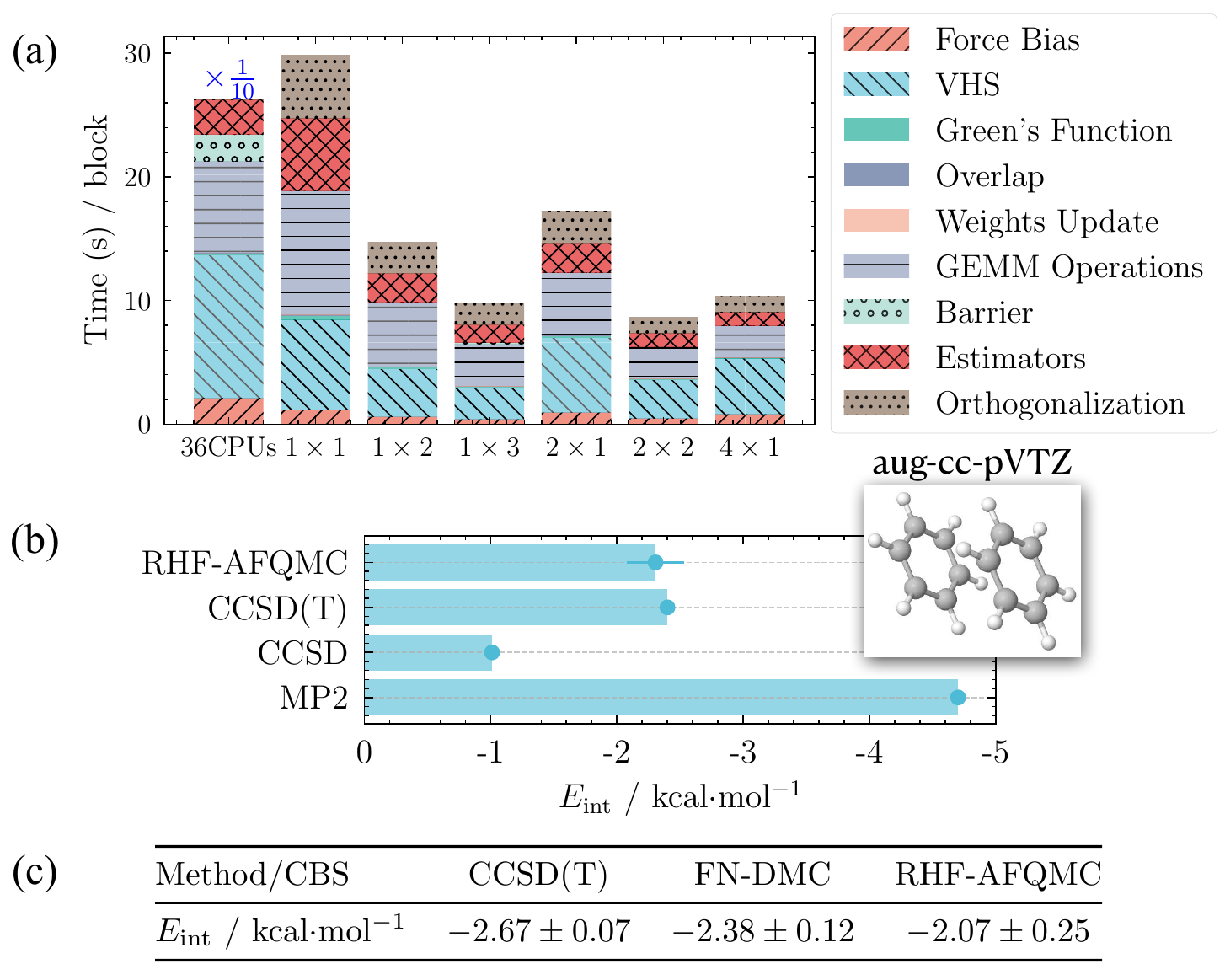}
    \caption{(a) Timing benchmarks for AFQMC calculations on the benzene dimer with the aug-cc-pVTZ basis employing 36 CPUs with shared memory and multiple A100 GPUs with the distributed Hamiltonian. The data were obtained by averaging 100 blocks, each containing 25 steps with a time step of 0.005$E_\textrm{h}^{-1}$. The interaction energy of the dimer computed from AFQMC is compared against other methods using the aug-cc-pVTZ basis in (b) and in the complete basis set limit obtained from a two-point basis extrapolation~\cite{halkier1998basis} from aug-cc-pVTZ and aug-cc-pVQZ calculations in (c). The MP2, CCSD, and CCSD(T) data using the aug-cc-pVTZ basis in (b) were obtained from Refs.~\citenum{gatech_bfdb_s22} and \citenum{burns2014appointing}. The CCSD(T) and FN-DMC data in (c) were extracted from Ref.~\citenum{al2021interactions}. 
    All AFQMC results were obtained with 1224 total walkers.
    \add{
    The CPU calculations employed Intel\textsuperscript{\textregistered} Xeon\textsuperscript{\textregistered} Platinum 8268 CPU @ 2.90GHz. The GPU calculations employed NVIDIA A100 GPU, accompanied with Intel\textsuperscript{\textregistered} Xeon\textsuperscript{\textregistered} Platinum 8358 CPU @ 2.60GHz.} 
    }
    \label{fig:benzene_PD}
\end{figure}

\add{We analyze computational wall time as a function of the configuration 
    (\texttt{nmembers}$\times$\texttt{ngroups}) in Fig.~\ref{fig:benzene_PD}(a), where we also compare the GPU implementation against the CPU setup.
    We denote the number of cards over which the Cholesky vectors are distributed as \texttt{nmembers}, and
    every \texttt{nmembers} cards form a group, as described in Fig.~\ref{fig:chunking_ring}(a).
    We denote the number of groups as \texttt{ngroups}.
    For benzene dimer with aug-cc-pVTZ, 
    a single A100 GPU card achieves performance around tenfold faster 
    than 36 CPU cores.
    Comparisons between configurations of $(1\times2)$ vs. $(2\times1)$ 
    and $(2\times2)$ vs. $(4\times1)$, with an equivalent total number of GPU cards, 
    demonstrated that configurations with a greater number 
    of \texttt{nmembers} were slightly slower due to the increased communication cost that scales linearly with \texttt{nmembers}. 
    We show how the communication within a group occurs in Fig.~\ref{fig:chunking_ring}(b).
    As shown in Fig.~\ref{fig:benzene_PD}(a), distributing the Hamiltonian across additional GPUs slightly slows 
    down the performance, predominantly during the \texttt{VHS} step, 
    which involves the construction of two-body propagators 
    (where HS stands for Hubbard-Stratonovich) during propagation.
    A $(2\times2)$ setup shows 2x faster speed compared to a $(2\times1)$ 
    configuration, reaffirming that the distributed memory approach retains good parallel efficiency.
    While communication overhead is an 
    inherent challenge in the distributed memory setup, this example shows the utility of our distributed-memory implementation. }

Using our GPU implementation, we calculated the interaction energy of the benzene dimer with counterpoise correction.~\cite{boys1970calculation} We compared AFQMC results with those from other methodologies, as shown in Fig.~\ref{fig:benzene_PD}(b)(c). While coupled cluster with singles and doubles (CCSD) underestimates the binding energy, second-order M{\o}ller-Plesset perturbation theory overestimates. Crucially, we observed that RHF-AFQMC is on par with CCSD with perturbative triples (CCSD(T)). 
\add{We used 1224 walkers for this calculation, which were found to have negligible population control biases.
In general, one should perform a convergence test to determine the number of walkers needed as is standard in QMC calculations~\cite{lee_twenty_2022}. We obtained more than 30,000 total blocks for both dimer and monomer calculations 
to compute the interaction energy with the error bars shown in Fig.~\ref{fig:benzene_PD}(b) and (c).
The raw data for all blocks can be obtained from the data repository in Section.~\ref{sec:data}.
Such a large number of blocks is generally unusual for AFQMC calculations,
in the rest of calculations shown in this paper, a few thousand blocks are sufficient
to achieve the desired accuracy.
}

\subsection{Support of complex Cholesky vectors}
Most calculations in quantum chemistry deal with real-valued integrals. 
However, in some cases, especially in the presence of external magnetic fields, when considering spin-orbit coupling or employing Bloch orbitals in solid-state calculations, the wavefunctions (and orbitals) can become complex-valued.~\cite{blaschke_cholesky_2022} This adds a layer of mathematical and computational complexity.
To handle these cases, \texttt{ipie} supports complex Cholesky vectors; the number of auxiliary fields is hence twice as many as the number of Cholesky vectors.
The implementation details are available in Ref.~\citenum{suewattana2007phaseless}.

\subsection{GPU accelerated MSD-AFQMC and timing benchmarks}\label{sec:msd_gpu}
ph-AFQMC with MSD trials is a powerful tool for systems with multi-reference characteristics, especially when the phaseless bias can be converged away.
Building on top of our successful CPU implementation~\cite{malone_ipie_2023} based on Wick's theorem,~\cite{mahajan2022selected} As detailed in Ref. \citenum{msdipie}, \texttt{ipie} has now enabled GPU-accelerated MSD-AFQMC implementations.
Here, we summarize some of the important features of our implementation and discuss applications of this feature.
Interested readers are referred to Ref. \citenum{msdipie} for more details.

Similar to our single determinant GPU implementation, \texttt{ipie} employs \texttt{cupy.einsum} via the \texttt{cuTENSOR} library to enhance matrix multiplication performance. Additionally, Wick's algorithm is adapted for GPU execution using custom CUDA kernels, effectively speeding up the computational hotspots in MSD-AFQMC, such as Green's function, overlap, and local energy computations.

\begin{figure}
    \centering
    \includegraphics[width=0.45\textwidth]{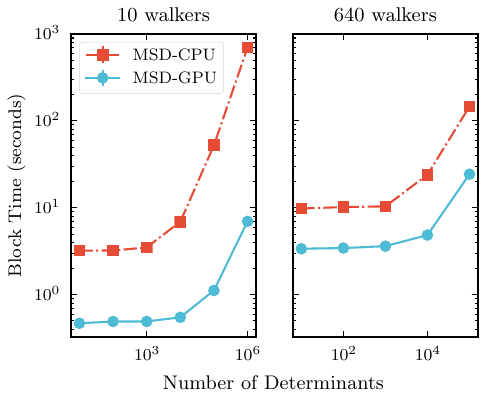}
    \caption{(a) Comparisons of the time cost per block for AFQMC calculations on [Cu$_2$O$_2$]$^{2+}$ using the BS1 basis ~\cite{malone_ipie_2023} with 10 walkers on a single CPU core and a single A100 GPU. 
    \add{The CPU calculations employed Intel\textsuperscript{\textregistered} Xeon\textsuperscript{\textregistered} Platinum 8336C CPU @ 2.30GHz}. (b) Time per block with 640 walkers on a single A100 against 32 CPU cores. Figure adapted from Ref.~\citenum{msdipie}.}
    \label{fig:msd_gpu}
\end{figure}

The GPU implementation of MSD-AFQMC is benchmarked on the example [Cu$_2$O$_2$]$^{2+}$ from \texttt{ipie}'s first release paper.~\cite{malone_ipie_2023}
In addition to employing a total of 10 walkers to directly compare with our previous results in Ref.~\citenum{malone_ipie_2023}, calculations using a total of 640 walkers were also performed to obtain a more realistic comparison.
With 10 walkers, the GPU implementation on a single NVIDIA A100 card is six times faster than using a single CPU core when the number of determinants in the trial is less than $10^3$.  As the number of determinants increases, the performance improvement is even more pronounced: tenfold with $10^4$ determinants and approximately 100-fold with $10^6$ determinants, as illustrated in Fig.~\ref{fig:msd_gpu}(a).
With 640 walkers, the GPU code on a single A100 card is compared against with 32 CPU cores.
As shown in Fig.~\ref{fig:msd_gpu}(b), a fourfold speedup is observed with less than $10^2$ determinants that increase to around sixfold with more determinants. 
The determinants were divided into chunks and computed sequentially to handle a larger number of determinants within the memory limits. This approach introduces an overhead that mostly affects the energy estimator, as reported in Ref.~\citenum{msdipie}.

\begin{figure}
    \centering
    \includegraphics[width=0.4\textwidth]{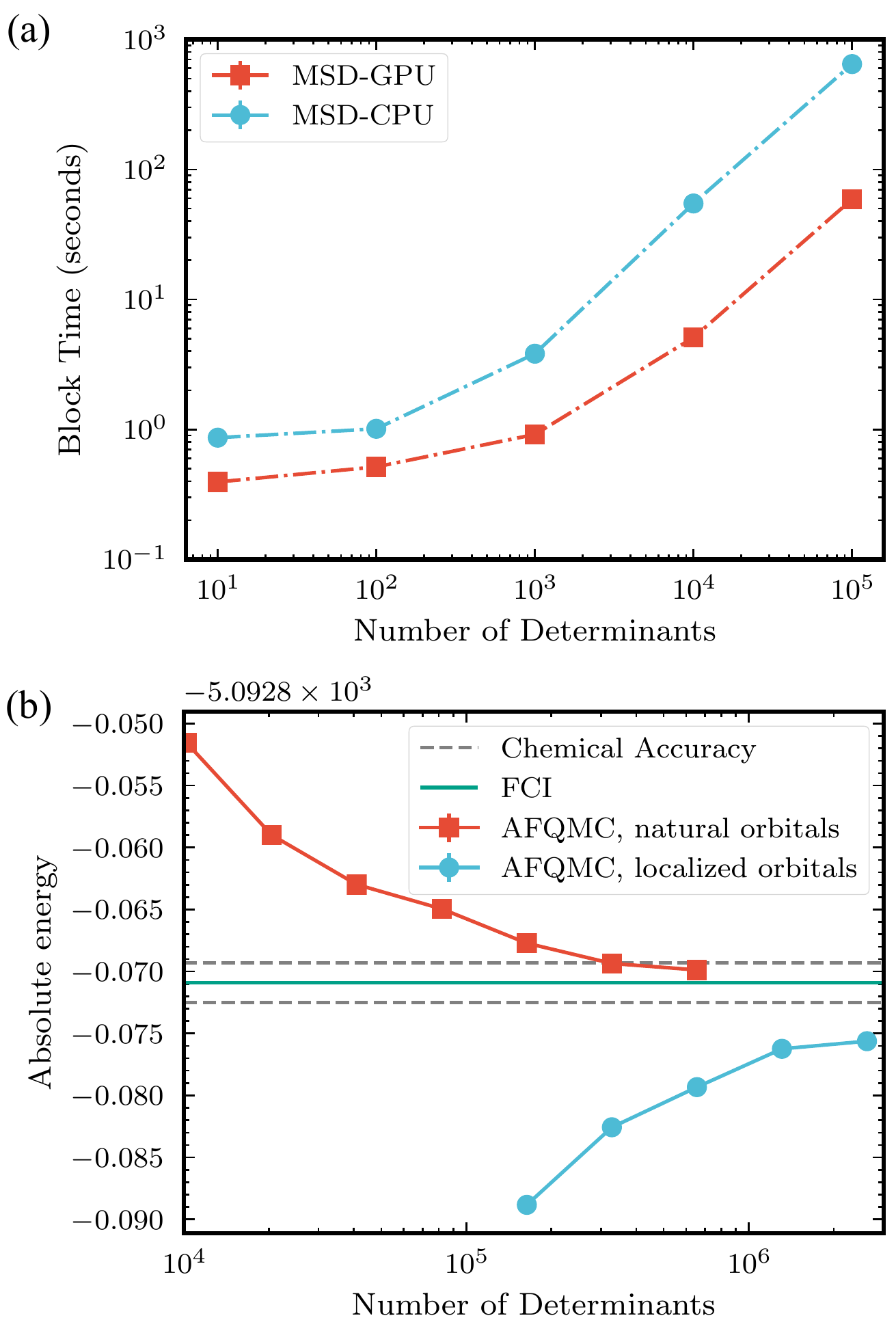}
    \caption{Time cost and absolute energy benchmarks on the [Fe$_2$S$_2$(SCH$_3$)$_4$]$^{2-}$ cluster. (a) The time cost per block with 640 walkers on 32 CPU cores and a single A100 GPU. \add{The CPU calculation employed Intel\textsuperscript{\textregistered} Xeon\textsuperscript{\textregistered} Platinum 8336C CPU @ 2.30GHz}. (b) Absolute energies are derived using localized atomic orbitals and natural orbitals. Figure adapted from Ref.~\citenum{msdipie}.}
    \label{fig:msd_gpu_fes}
\end{figure}

The timing and absolute energy benchmarks are further benchmarked on the (20o, 30e) active space of the [Fe$_2$S$_2$(SCH$_3$)$_4$]$^{2-}$ cluster,~\cite{msdipie} as shown in Fig.~\ref{fig:msd_gpu_fes}.
In contrast to the 108-orbital calculation of [Cu$_2$O$_2$]$^{2+}$, a merely two-fold speedup is observed using a single A100 card compared to 32 CPUs. This speedup increases to almost tenfold when the number of determinants exceeds $10^4$.
As shown in Fig.~\ref{fig:msd_gpu_fes}(b), another observation is using natural orbitals for AFQMC calculation leads to chemical accuracy with around 3$\times10^5$ determinants in the trial.
In contrast, reaching chemical accuracy is more challenging when using localized orbitals employed for DMRG calculations.~\cite{li2017spin}
These observations highlight the critical need to select an appropriate set of orbitals for AFQMC calculations involving strongly correlated systems. 
Additionally, it is noteworthy that the energies obtained using localized orbitals with a limited number of determinants are initially lower than the FCI energy, and they tend to converge upward toward the FCI reference as the number of determinants increases.

\subsection{Free projection AFQMC}
Free projection (fp-) AFQMC is a numerically exact method for calculating the eigenvalues of a Hamiltonian. While the more commonly used ph-AFQMC variant employs the phaseless constraint to control the sign problem, fp-AFQMC attempts to sample the ground state energy brute-force with exponential-scaling sample complexity. Despite this scaling, it is possible to perform relatively large active space calculations in practice using accurate trial states. 

In fp-AFQMC, the ground-state energy is estimated by
\begin{equation}
   E(\tau) = \frac{\langle\psi_l|\hat{H} \mathrm{e}^{-\tau \hat{H}}|\psi_r\rangle}{\langle\psi_l|\mathrm{e}^{-\tau \hat{H}}|\psi_r\rangle} = \frac{\int d\mathbf{x} \ p(\mathbf{x})\langle\psi_l|\hat{H} \hat{B}(\mathbf{x},\Delta\tau)|\psi_r\rangle}{\int d\mathbf{x} \ p(\mathbf{x})\langle\psi_l|\hat{B}(\mathbf{x},\Delta\tau)|\psi_r\rangle},
\end{equation}
where \(\ket{\psi_l}\) and \(\ket{\psi_r}\) are left and right trial states, respectively; \(\hat{B}(\mathbf{x},\Delta\tau)\) is defined in Eq.~\ref{eq:Bprop}; and \(\tau\) is the imaginary time. Analogously to the ph-AFQMC propagator sampling, the auxiliary fields \(\mathbf{x}\) are sampled from the Gaussian distribution \(p(\mathbf{x})\). Unlike ph-AFQMC, we do not use the force bias to perform importance sampling. Instead, only mean-field subtraction is used. We also do not employ population control.

A judicious choice of \(\ket{\psi_l}\) and \(\ket{\psi_r}\) offers two advantages. First, more accurate trial states (\textit{eg.}, selected CI) reduce the imaginary time required to project the ground state energy to a given accuracy, thereby reducing the noise in the energy estimate.~\cite{mahajan2021taming,mahajan2022selected} 
Integrating the GPU-accelerated MSD code~\cite{msdipie} will also significantly accelerate the fp-AFQMC calculations with MSD trials.
Second, since energies are measured using the state \(\ket{\psi_l}\), the closer this state is to the ground state, the smaller the variance in the energy estimate due to the zero variance principle.~\cite{assaraf_zero-variance_1999}  
Furthermore, one can use a CCSD wavefunction as the initial state \(\ket{\psi_r}\), which reduces the projection time. CCSD wavefunction is employed by performing a Hubbard–Stratonovich transformation on the exponential of the cluster operator.~\cite{mahajan2021taming}
Using trial states belonging to specific symmetry sectors also allows one to target the lowest energy states in the corresponding sectors.

\begin{figure}[h]
	\centering
	\includegraphics[width=0.45\textwidth]{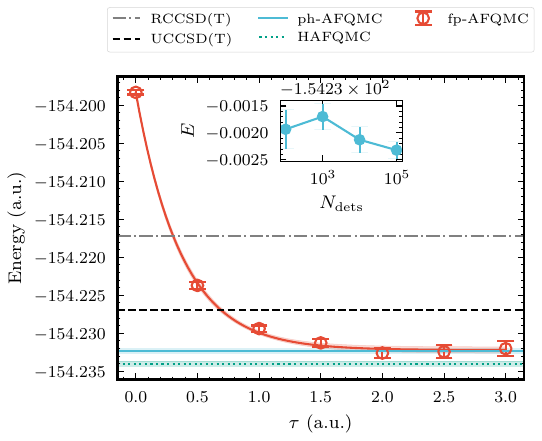}
	\caption{Convergence of the fp-AFQMC energy for the cyclobutadiene transition state in the cc-pVDZ basis set with projection time. The inset shows the convergence of ph-AFQMC energy with respect to the number of determinants in the trial.}\label{fig:cbd}
\end{figure}

As an illustrative example, we consider the \(D_{4h}\) symmetric transition state of cyclobutadiene. This state has a biradical character, which makes it a challenging problem for single reference electronic structure methods, including spin-restricted and spin-unrestricted CCSD(T) (RCCSD(T) and UCCSD(T)). We performed fp-AFQMC calculations on this system using an MSD trial, obtained via heat-bath configuration interaction (HCI), as \(|\psi_l\rangle\) and the spin-restricted CCSD (RCCSD) state as \(|\psi_r\rangle\). The HCI state was obtained from an HCI calculation with a crude \(\epsilon_1 = 10^{-4}\) in the full space except for four frozen HF orbitals. These orbitals were kept frozen in all correlated calculations. The geometry was taken from Ref.~\citenum{mahajan2021taming}, which also reported fp-AFQMC energies for this system. As additional validation of our results, we converged the ph-AFQMC energy with respect to the number of determinants in the trial HCI state; we expect this energy to be nearly exact. Results are shown in Fig.~\ref{fig:cbd}. 

The fp-AFQMC energy converges to the converged ph-AFQMC energy within statistical error bars. 
Our results suggest that hybrid (H)-AFQMC energies reported in Ref.~\citenum{chen2023hybrid} are likely biased as they are too much lower than our converged ph-AFQMC and fp-AFQMC energies.
RCCSD(T) is about 13.5 m$E_\textrm{h}$ higher than ph-AFQMC, while UCCSD(T) is closer but still about 4.5 m$E_\textrm{h}$ higher. The differences in the barrier height largely come from transition state energies because all these methods work well for the equilibrium \(D_{2h}\) geometry.

\subsection{AFQMC beyond the ground-state electronic structure energy}
The preceding discussions concentrate on the ground-state 
\textit{ab initio} electronic structure energy calculations. 
Apart from this, \texttt{ipie} now also accommodates other types of 
calculations within the framework of AFQMC, such as property 
calculations,~\cite{mahajan2023response} 
finite temperature calculations~\cite{PhysRevB.99.045108,lee2021phaseless} and 
ground-state calculations of electron--phonon coupled model systems.~\cite{lee2021constrained}
The code structures for these different features mirror those of the zero temperature \textit{ab initio} electronic structure implementation.
\add{We also note that \texttt{ipie}'s legacy code also supports
the ground state and finite-temperature calculations of other model systems such as the Hubbard model, which
we do not discuss in this paper.}

\subsubsection{Electronic structure at finite temperatures}
An extension of the phaseless AFQMC method at zero temperature, finite-temperature AFQMC (FT-AFQMC), was developed to study systems at finite temperatures.~\cite{lee2021phaseless} It is customary (though not necessary~\cite{Shen2020Nov}) to work in the grand canonical ensemble described by the temperature $T$, volume $V$, and chemical potential $\mu$. The central quantities of interest are thermal expectation values computed from the partition function
\begin{align}
    \Xi = \mathrm{Tr} \left[ \mathrm{e}^{-\beta(\hat{H} - \mu \hat{N})} \right], \label{trace}
\end{align}
where $\hat{N}$ is the total number operator and $\mu$ is the chemical potential. The Boltzmann factor in Eq.~\eqref{trace} can be interpreted as a propagator in imaginary time $\tau = \beta$. 

Analogously to the zero temperature case, $\tau$ is first discretized into $l$ intervals of length $\Delta \tau = \tau / l$,
\begin{align}
    \Xi = \mathrm{Tr} \left[ \mathrm{e}^{-\beta(\hat{H} - \mu \hat{N})} \right] = \mathrm{Tr} \left[ \lim_{l \to \infty} \prod_{k=1}^{k=l} \mathrm{e}^{-\Delta \tau(\hat{H} - \mu \hat{N})} \right],
\end{align}
the short-time propagator is then Trotter decomposed as
\begin{align}
    \mathrm{e}^{-\Delta \tau(\hat{H} - \mu \hat{N})} \simeq \mathrm{e}^{-\frac{\Delta \tau}{2} (\hat{v}_0 - \mu \hat{N})} \mathrm{e}^{\frac{\Delta \tau}{2} \sum_{\gamma} \hat{v}_\gamma^2} \mathrm{e}^{-\frac{\Delta \tau}{2} (\hat{v}_0 - \mu \hat{N})},
\end{align}
and the application of the Hubbard--Stratonovich transformation gives
\begin{align}
    \mathrm{e}^{-\Delta \tau(\hat{H} - \mu \hat{N})}
    &\simeq \int d\mathbf{x} \ p(\mathbf{x}) \hat{B}(\mathbf{x}, \Delta \tau, \mu),
\end{align}
where the one-body propagator $\hat{B}$ is now also function of $\mu$:
\begin{equation}
    \hat{B}(\mathbf{x}, \Delta\tau, \mu) = \mathrm{e}^{-\frac{\Delta\tau}{2} (\hat{v}_0 - \mu \hat{N})} \mathrm{e}^{-\sqrt{\Delta \tau} \mathbf{x}\cdot \hat{\mathbf{v}}}\mathrm{e}^{-\frac{\Delta\tau}{2} (\hat{v}_0 - \mu \hat{N})}.\label{eq:Bprop_thermal}
\end{equation}
The grand canonical partition function is thus evaluated as
\begin{align}
    \Xi &= \mathrm{Tr} \left[ \mathrm{e}^{-\beta(\hat{H} - \mu \hat{N})} \right] \nonumber \\
    &= \int d\mathbf{x}_1 \cdots d\mathbf{x}_l \ \underbrace{p(\mathbf{x}_1) \cdots p(\mathbf{x}_l)}_{p(\mathbf{x}_1, \dots, \mathbf{x}_l)} \ \mathrm{Tr} \left[ \prod_{k=1}^l \hat{B}(\mathbf{x}_k, \Delta \tau, \mu) \right], \label{gc_partition_tr}
\end{align}
with $p(\mathbf{x}_1, \dots, \mathbf{x}_l)$ being the probability of sampling a specific path designated by auxiliary fields $\mathbf{x}_1, \dots, \mathbf{x}_l$. Furthermore, the trace in Eq.~\eqref{gc_partition_tr} can be written in terms of a determinant,~\cite{Blankenbecler_1981,Hirsch_1985,Santos_2003} which finally yields
\begin{align}
    \Xi &= \int d\mathbf{x}_1 \cdots d\mathbf{x}_l \ p(\mathbf{x}_1, \dots, \mathbf{x}_l) \ \mathrm{det} \left[ \mathbf{I} + \prod_{k=1}^l \mathbf{B}(\mathbf{x}_k, \Delta \tau, \mu) \right], \label{gc_partition_det}
\end{align}
where $\mathbf{I}$ is the identity matrix and $\mathbf{B}$ is a matrix representation of $\hat{B}$ in a single-particle basis. 
It should be noted that the partition function itself is not explicitly calculated--expectation values derived from it are. For some generic observable $\hat{O}$, its expectation value is
\begin{align}
    \braket{\hat{O}} &= \frac{1}{\Xi} \mathrm{Tr} \left[ \mathrm{e}^{-\beta (\hat{H} - \mu \hat{N})} \hat{O}\right] \label{thermal_avg} \\
    &= \int d\mathbf{X} \
    \frac{p(\mathbf{X}) \ \mathrm{Tr} \left[A(\mathbf{X}, \Delta \tau, \mu)\right]}{\int d\mathbf{Y} \ p(\mathbf{Y}) \ \mathrm{Tr} \left[A(\mathbf{Y}, \Delta \tau, \mu)\right]} \times \\
    & \frac{\mathrm{Tr} \left[ A(\mathbf{X}, \Delta \tau, \mu) \hat{O} \right]}{\mathrm{Tr} \left[A(\mathbf{X}, \Delta \tau, \mu)\right]}, \label{thermal_avg_is}
\end{align}
where we introduced the shorthand $\mathbf{X}$ for the set of auxiliary fields along an imaginary path $\left\{\mathbf{x}_1, \cdots, \mathbf{x}_l\right\}$, and 
\begin{align}
    A(\mathbf{X}, \Delta \tau, \mu) &= \prod_{k=1}^l \hat{B}(\mathbf{x}_k, \Delta \tau, \mu).
\end{align}
Rewriting Eq.~\eqref{thermal_avg} in the form Eq.~\eqref{thermal_avg_is} allows $\braket{\hat{O}}$ to be estimated through an importance sampling procedure. The field configurations $\mathbf{X}$ are obtained via Monte Carlo sampling from the modified probability distribution $\tilde{p}(\mathbf{X}) = p(\mathbf{X}) \ \mathrm{Tr} \left[A(\mathbf{X}, \Delta \tau, \mu)\right]$, while the computed random variables are the \textit{local} expectation values
\begin{align}
    O_L(\mathbf{X}, \Delta \tau, \mu) = \frac{\mathrm{Tr} \left[ A(\mathbf{X}, \Delta \tau, \mu) \hat{O} \right]}{\mathrm{Tr} \left[A(\mathbf{X}, \Delta \tau, \mu)\right]}.
\end{align}

\begin{figure}[H]
    \centering
    \includegraphics[width=0.45\textwidth]{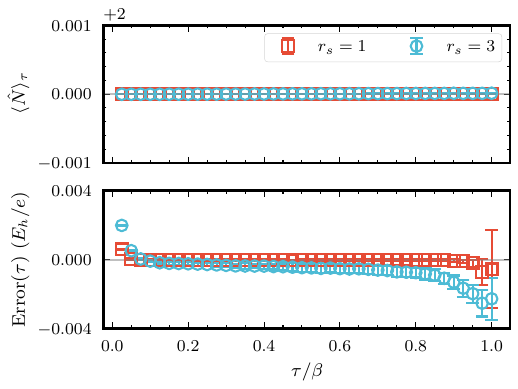}
    \caption{Imaginary time trace from FT-AFQMC of the uniform electron gas model with 2 electrons. The top panel depicts the average electron number at imaginary time $\tau$, $\braket{\hat{N}}_{\tau}$, while the bottom panel shows the error in the total energy compared to FCI, \textit{i.e.} $E_\text{FT-AFQMC}(\tau) - E_\text{FCI}$. 
    These results reproduce Fig. 3 in Ref.~\citenum{lee2021phaseless}.}
    \label{fig:ft_afqmc_im_trace}
\end{figure}

Similar to the zero temperature case, the importance sampling is implemented by initializing a set of walkers in the space of auxiliary fields with weights $w_i$ and propagating them in imaginary time. We therefore evaluate Eq.~\eqref{thermal_avg_is} in practice as
\begin{align}
    \braket{\hat{O}} = \frac{\sum_i w_i O_L(\mathbf{X}_i)}{\sum_i w_i}.
\end{align}
An example of the imaginary time trace for observables is provided in Fig. \ref{fig:ft_afqmc_im_trace}, which reproduces Fig. 3 in Ref.~\citenum{lee2021phaseless}.

\subsubsection{Electrons coupled to phonons}

Utilizing the tools provided in \ipie{}, one may write a new projector Monte Carlo method to calculate the ground state of electrons coupled to phonons (i.e., lattice vibration).\cite{lee2021constrained} 
\begin{equation}
\hat{H} = \underbrace{\sum_{pq}h_{pq}a_p^\dagger a_q}_{\hat{H}_\text{el}}
+ \underbrace{\sum_\nu \omega_\nu b_\nu^\dagger b_\nu}_{\hat{H}_\text{ph}}
+
\underbrace{\sum_{pq\nu}g_{pq\nu} a_p^\dagger a_q (b_\nu + b_\nu^\dagger)}_{\hat{H}_\text{el-ph}}
\label{eq:ephham}
\end{equation}
where $b^{(\dagger)}$ represents the bosonic annihilation (and creation) operator, 
the first term represents the electronic band term, the second term represents the phonon band term, and the third term represents the coupling between electrons and phonons.


The corresponding ground-state projector is obtained from trotterizing the imaginary time propagator corresponding to Eq.~\eqref{eq:ephham},
\begin{align} \label{eq:holstein_propagator}\nonumber
    \mathrm{e}^{-\Delta\tau \hat{H}} &= \mathrm{e}^{-\frac{\Delta\tau}{2} \hat{H}_{\mathrm{el}}} \mathrm{e}^{-\frac{\Delta\tau}{2} \hat{H}_{\mathrm{ph}}} \mathrm{e}^{-\Delta\tau \hat{H}_{\mathrm{el-ph}}} \mathrm{e}^{-\frac{\Delta\tau}{2} \hat{H}_{\mathrm{ph}}} \mathrm{e}^{-\frac{\Delta\tau}{2} \hat{H}_{\mathrm{el}}}\\ 
    &+ \mathcal{O}(\Delta\tau^3)
\end{align}
In our method, we choose walkers of the form $|\psi_\mathrm{w}\rangle \otimes |\mathbf{X}_\mathrm{w}\rangle \equiv |\psi_{\mathrm{w}}(\tau),\mathbf{X}_\mathrm{w}(\tau)\rangle$, with $|\psi_\mathrm{w}\rangle$ being a single determinant and $|\mathbf{X}_\mathrm{w}\rangle$ being the coordinates for phonon displacements. 
Using the Monte Carlo sampling, we can work in the position space of phonons without invoking any boson number truncation.
This strategy differs from other standard Monte Carlo approaches in this area.~\cite{macridin_phonons, berciu_polaron}

\begin{figure}
    \centering
    \includegraphics[width=0.5\textwidth]{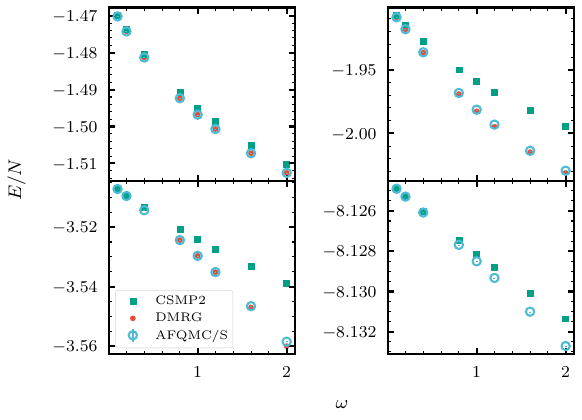}
    \caption{Reproduction of Fig.~3 in Ref.~\citenum{lee2021constrained}. Results are shown for a 20-site Holstein model at half-filling. We used a coherent state trial to obtain the AFQMC/S results. CSMP2 data points correspond to perturbation theory-based computations based on molecular orbitals obtained from SCF calculations for a Holstein Fock operator.\cite{lee2021constrained} Panels (a)-(d) show scans along phonon frequencies $\omega$ for $\lambda$ equals (a) 0.1, (b) 0.3, (c) 0.8, and (d) 2.}
    \label{fig:holstein_20e_20o}
\end{figure}

With importance sampling via a trial wavefunction, $|\Psi_\text{T}\rangle$, our global wavefunction is a weighted linear combination of walker vibronic wavefunctions,
\begin{equation}
    |\Psi(\tau)\rangle = \sum_{\mathrm{w}=1}^{N_\text{walkers}} 
    w_{\mathrm{w}}(\tau) \: 
    \frac{|\psi_{\mathrm{w}}(\tau),\mathbf{X}_\mathrm{w}(\tau)\rangle}{\langle \Psi_\text{T}|\psi_{\mathrm{w}}(\tau),\mathbf{X}_\mathrm{w}(\tau)\rangle}.
\end{equation}
The propagation of the bosonic degrees of freedom under $\hat{H}_{\mathrm{ph}}$ is performed via a diffusion Monte Carlo algorithm.~\cite{Kalos1974ImportanceDMC}
With the walker representation, the propagation with $\hat{H}_\text{el}$ and $\hat{H}_\text{el-ph}$ is straightforward by exploiting the Thouless theorem.
More details can be found in Ref.~\citenum{lee2021constrained}.
As an example calculation, we picked the one-dimensional Holstein model under a periodic boundary condition,~\cite{holstein} which involves $h_{pq} = -t (\delta_{p+1,q} + \delta_{p,q+1})$, $\omega_\nu = \omega$, and $g_{pq\nu} = g \delta_{pq}\delta_{p\nu}$.
For 20-site at half-filling, we considered four different unitless electron--phonon coupling strengths $\lambda=\frac{g^2}{2t\omega}$, which are displayed in Fig.~\ref{fig:holstein_20e_20o}.
\add{It is evident from Fig.~\ref{fig:holstein_20e_20o} that the coherent state trial is not an optimal choice for the intermediate coupling regime, where $2 t \lambda < \omega$. By employing more sophisticated trial wavefunctions, we can obtain more reliable results also for the intermediate coupling regime.\cite{lee2021constrained}
The computational bottleneck of our QMC algorithm is the evolution under the el--ph projection operator, generally scaling as $\mathcal{O}(N^3)$.}

\subsubsection{Automatic differentiatiable AFQMC}
The computation of observables that do not commute with $\hat{H}$ poses additional challenges in projector Monte Carlo, such as AFQMC.
A recently proposed approach by Mahajan \textit{et. al.}\cite{mahajan2023response} aims to compute the response estimator (Eq.~\eqref{resp}), which, though still not exact due to the discontinuity in the distribution, is accurate enough to give reliable results: 
\begin{equation}
    \langle \hat{O} \rangle_{\text{response}} = \frac{\mathrm{d} E_{\text{AFQMC}}(\lambda)}{\mathrm{d} \lambda}\Bigg|_{\lambda = 0}.
    \label{resp}
\end{equation}
Here $\lambda$ is a parameter for the perturbed Hamiltonian: $\hat{H}(\lambda) = \hat{H} + \lambda \hat{O}$, and AFQMC energy $E_{\text{AFQMC}}(\lambda)$ is given by the Monte Carlo estimator
\begin{equation}
    E_{\text{AFQMC}}(\lambda) = \frac{\sum_i w_i(\lambda) \frac{\langle\Psi_T(\lambda)|\hat{H}(\lambda)| \psi_i(\lambda)\rangle}{\langle\Psi_T(\lambda) | \psi_i(\lambda)\rangle}}{\sum_i w_i(\lambda)},
    \label{afqmc_energy}
\end{equation}
where $w_i(\lambda)$ is the weight for the $i$-th walker, and $|\psi_i(\lambda)\rangle$ is the $i$-th walker state at coupling strength $\lambda$.

We implemented this scheme for computing observables using the Automatic Differentiation (AD) functionality of \texttt{PyTorch}\cite{NEURIPS2019_9015} as an add-on within \texttt{ipie}. 
It is noteworthy that storing the computation graph for an entire AFQMC run requires substantial memory. In practice, we use the concept of AD blocks.\cite{mahajan2023response} 
We only track the computation graph within an AD block; there is no connection between different AD blocks. 
Using this approach, we manage memory costs by adjusting block size. 
We also use the gradient checkpointing technique\cite{chen2016training} to reduce the memory cost further. 

For relatively large systems, the AFQMC calculation is parallelized using MPI to distribute walkers over MPI tasks. The differentiation of the AFQMC global energy estimator is not embarrassingly parallel since walkers will mix between different MPI ranks via population control. 
The AD implementation in \ipie{} does not support communication between MPI tasks.
Therefore, we perform ``local'' population control within an MPI rank. Because the local population control does not mix the walkers between different ranks, it is valid to regard those estimates as independent samples. We thus perform block analysis on those samples to obtain the final result.
Effectively, this amounts to running very low walker population simulations (50 walkers per task here), which is only practically possible for small system sizes where population control is not a concern. For intermediate system sizes, one could imagine using large memory nodes and OpenMP threading or using single GPUs with a sufficient amount of memory.

Here, we present the results of AD-AFQMC calculations on various molecular systems. The molecular integrals are obtained by \texttt{PySCF},\cite{sun2018pyscf} and the modified Cholesky decomposition is performed with \texttt{ipie}, with a threshold of $10^{-5}$. We used a time step of 0.01 a.u., and periodic reorthogonalization of walkers is performed every 5 steps for all calculations. We used the restricted Hartree Fock (RHF) trial for all systems. For accurate statistical analysis, we perform block analysis for $
\geq 200$ gradient samples for each calculation. 
All AFQMC calculations are performed using the frozen-core approximation.

We benchmarked our implementation on various small molecules using aug-cc-pVTZ~\cite{dunning1989gaussian} basis set and compared the results to Ref.~\citenum{mahajan2023response} in \cref{tab:adafqmc}. All dipole moments align strongly with the AD-AFQMC results reported in Ref.~\citenum{mahajan2023response}. Furthermore, except for CO, AD-AFQMC matches experimental values better than RCCSD and exhibits accuracy comparable to RCCSD(T). This supports the widely held view that AFQMC's accuracy falls between CCSD and CCSD(T). 

\begin{table*}
    \caption{
    \label{tab:adafqmc}
    Comparison of AD-AFQMC dipole moment (in a.u.) of various molecules at equilibrium geometry with the implementation in Ref.~\citenum{mahajan2023response} and other quantum chemistry methods. The data using RCCSD and RCCSD(T) are also extracted from Ref.~\citenum{mahajan2023response}.}
    \begin{ruledtabular}
    \begin{tabular}{cccccc}
        Molecule & This work & AD-AFQMC in Ref.~\citenum{mahajan2023response} & RCCSD  & RCCSD(T) & Experiment \\
        \hline
        \ce{H2O}& 0.723(2)    & 0.720(2)       & 0.7335 & 0.7247   & 0.730\cite{shostak1991dipole}   \\
        \ce{NH3}      & 0.592(2)    & 0.592(2)       & 0.6015 & 0.5938   & 0.581(1)\cite{shimizu1970stark}   \\
        CO       & 0.022(4) & 0.019(4)       & 0.0199 & 0.0429   & 0.048(1)\cite{MUENTER1975490}   \\
        HCl      & 0.428(1)    & 0.429(1)       & 0.4318 & 0.4273   & 0.430\cite{139951}      \\
        HBr      & 0.332(2)    & 0.329(2)       & 0.3289 & 0.3245   & 0.325\cite{139951}
    \end{tabular}
    \end{ruledtabular}
    \end{table*}

\subsection{Enhanced integration testing and no-MPI mode}
As \texttt{ipie} expands its functionalities, robust testing workflows become crucial.
\texttt{ipie} supports a new integration testing framework that enhances the package's robustness and adaptability and significantly contributes to its reliability and ease of use for the end-users.
The GitHub continuous integration (CI) workflow automatically tests new pull requests, ensuring that every change to the codebase does not break existing functionalities. The workflow encompasses a comprehensive test suite and executes various linting and code formatting checks before running serial and parallel unit tests and integration tests.

Recognizing the diverse computational environments in which \texttt{ipie} might be deployed, we introduced the no-MPI mode. This mode is designed for situations where MPI is unavailable, or its use is not desired, offering greater flexibility for users and developers. The CI workflow includes a job that tests \texttt{ipie}'s functionality without the MPI dependency.
This mode is particularly beneficial for users who wish to perform quantum Monte Carlo simulations on personal computing setups or in environments where setting up MPI is challenging.

\section{Interfaces to external packages}\label{sec:interfaces}
\subsection{Dice interface and SHCI-AFQMC}
\texttt{ipie} includes utilities for converting the output from Dice,~\cite{sharma2017semistochastic} a package that employs semistochastic heat bath configuration interaction (SHCI)~\cite{holmes2016heat,sharma2017semistochastic} as the complete active space (CAS) solver.
\begin{listing}[H]
\begin{mycode}
python -u /path-to-ipie/tools/extract_dice.py --dice-wfn /path-to-Dice-output/dets.bin --sort --verbose
\end{mycode}
\caption{Convert Dice output to the MSD trial in \texttt{ipie}.}
\label{lst:dice_convert}
\end{listing}
\noindent which produces the \texttt{wfn.h5} file containing the coefficients and the indices of occupied orthogonal orbitals for spin-up and spin-down sectors that follows the block-formatted ($\alpha\alpha\cdots\beta\beta\cdots$) orbital convention in \texttt{ipie}.
The SHCI calculation generates a good MSD trial for challenging systems.~\cite{malone_ipie_2023,mahajan2022selected}
However, it is not a black-box approach and often requires careful handling, such as selecting CAS and using natural orbitals, as discussed in Section~\ref{sec:msd_gpu}.

These detailed discussions are beyond the scope of this article, and we briefly mention the procedures in the semi-blackbox example provided within \texttt{ipie}:~\cite{malone_ipie_2023}
\begin{enumerate}
    \item Initiate a preliminary rough SHCI calculation in an extensive active space for the system under study.
    \item Derive the SHCI one-electron reduced density matrix (1-RDM), extracting the resultant natural orbital occupation number (NOON) and natural orbitals.
    \item Define an active space criterion based on a predetermined NOON threshold.
    \item Adjust the orbitals through rotation, aligning them with the unitary transformation specified by the natural orbitals.
    \item Execute a refined SHCI self-consistent field calculation within the new active space determined in the previous step.
    \item Use this MSD trial in AFQMC.
\end{enumerate}
This strategy ensures the trial wavefunction encapsulates static correlations within the active space via the MSD trial. At the same time, AFQMC incorporates the residual dynamic correlations. 
This procedure is folded into the factory utility method \texttt{ipie.utils.from\_dice.build\_driver\_from\_shciscf}.
\subsection{TREXIO support and CIPSI-AFQMC}
The \trexio{} library and file format have been developed to offer a robust and efficient solution for storing and exchanging wavefunction parameters and matrix elements.\cite{trexio}
This library supports bindings in several programming languages, including Python, and can be conveniently installed via the \texttt{pip} package manager.

The compatibility of \texttt{ipie} with the \trexio{} format facilitates its integration with various software packages.
Specifically, it allows \texttt{ipie} to utilize trial wavefunctions produced by Quantum Package,\cite{qp2} along with the associated one-electron integrals and Cholesky-decomposed electron repulsion integrals. This interface allowed us to check if AFQMC could complement configuration interaction using perturbative selection done iteratively (CIPSI) calculations to improve full configuration interaction (FCI) energy estimates of large systems.

Typically, to estimate the FCI energy from a CIPSI calculation, one extrapolates to zero the variational energy, $E_{\text{trial}}$, as a function of the renormalized second-order perturbative correction, $E_{\text{rPT2}}$. \cite{qp2}
However, the AFQMC energy, $E_{\text{AFQMC}}$, is anticipated to provide a closer approximation to the FCI energy than the sum of $E_{\text{trial}}$ and $E_{\text{rPT2}}$.
This expectation is particularly relevant for systems with large $E_{\text{rPT2}}$ corrections.

\newcommand{\tabc}[1]{\multicolumn{1}{c}{#1}}

\begin{figure}
    \centering
    \includegraphics[width=0.45\textwidth]{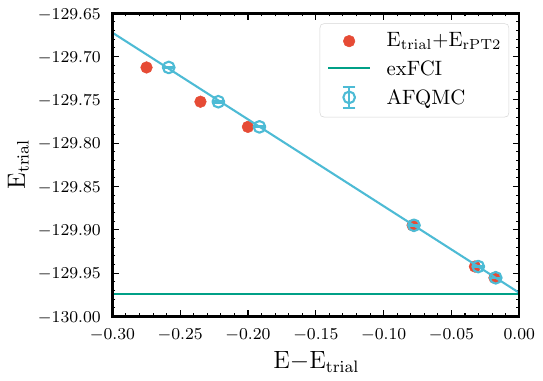}
    \caption{Energy of the trial wavefunction ($E_{\text{trial}}$) as a function of $\Delta E$, where $\Delta E = E_{\text{rPT2}}$ for CIPSI calculations and $\Delta E = E_{\text{AFQMC}} - E_{\text{trial}}$ for AFQMC. exFCI is the extrapolated FCI energy obtained from CIPSI calculations.}
    \label{fig:hno}
\end{figure}

\begin{table*}
\caption{
\label{tab:cipsi}
Renormalized PT2 correction $E_\text{rPT2}$, Variational energy $E_\text{var}$, and AFQMC energy $E_\text{AFQMC}$ for the nitroxyl and benzene molecules as functions of the number of determinants in the variational space.}
\begin{ruledtabular}
\begin{tabular}{ldddd}
System, Basis & \tabc{$N_\text{det}$} &  \tabc{$E_\text{trial}$} & \tabc{$E_\text{trial}+E_\text{rPT2}$} &     \tabc{$E_\text{AFQMC}$} \\
\hline

\ce{HNO}, 6-31G &    1  & -129.712554  & -129.987361   & -129.971(1)  \\
                &    2  & -129.752113  & -129.987158   & -129.974(1)  \\
                &   41  & -129.781149  & -129.981219   & -129.9727(6) \\
                &  748  & -129.894775  & -129.973405   & -129.9724(2) \\
                & 2838  & -129.942391  & -129.975057   & -129.9726(1) \\
                & 14201 & -129.955368  & -129.973620   & -129.9727(1) \\
\hline
\ce{C6H6}, cc-pVDZ &      1 & -230.7204904 & -231.39330 & -231.5887(7) \\
                   &     23 & -230.7615816 & -231.42167 & -231.5866(7) \\
                   &    828 & -230.8687072 & -231.44602 & -231.5864(6) \\
                   &  15690 & -231.1300087 & -231.49330 & -231.5848(4) \\
                   & 109869 & -231.3152399 & -231.53022 & -231.5842(4) \\
\end{tabular}
\end{ruledtabular}
\end{table*}

We conducted AFQMC calculations on the nitroxyl and benzene molecules using CIPSI trial wavefunctions of increasing sizes.
The results of these calculations are detailed in Table~\ref{tab:cipsi}.
Supporting our expectation, Fig.~\ref{fig:hno} demonstrates that the AFQMC corrections align the data points along a straight line, \cite{burton2024rationale} validating the hypothesis that AFQMC energies are more reliable for the extrapolation towards the FCI value, especially in cases with large rPT2 corrections.
Employing calculations with comparatively small wavefunctions for benzene, as shown in Table~\ref{tab:cipsi}, a three-point linear extrapolation based on the rPT2 correction yields a correlation energy of \SI{-858.6}{\milli\hartree}. In contrast, extrapolation using AFQMC energies results in a correlation energy of \SI{-862.2}{\milli\hartree}. This latter value is significantly closer to the correlation energy of \SI{-863.4}{\milli\hartree} achieved through CIPSI with 167 million determinants.\cite{loos_2020}

These preliminary calculations enabled by the \trexio{} interface illustrate
that the integration of AFQMC with CIPSI emerges as a promising methodology for
estimating the FCI energy of systems larger than those currently feasible.
\subsection{FQE interfaces}
The Fermionic quantum emulator (FQE)~\cite{fqe_2021,*rubin2021fermionic} is a lightweight fermionic circuit simulator, which is particularly useful in quantum computing where it aids in the development and testing of quantum algorithms tailored for fermionic systems. 
\texttt{ipie} provides the conversion between the \texttt{ipie}'s MSD wavefunction and the FQE wavefunction.

\section{Conclusions and outlooks}\label{sec:conclusion}
This paper summarized the improvements and new features added in \texttt{ipie} since its original release.\cite{malone_ipie_2023} 
These improvements enhance modularity and computational efficiency and offer intuitive user-end APIs. New features and interfaces aim to expand a broader spectrum of AFQMC calculations in quantum chemistry. 

We summarize the key features we highlighted in this manuscript:
\begin{enumerate}
    \item {\it Distributed Hamiltonians to remove the memory bottleneck.} We demonstrated \texttt{ipie}'s capacity for studying large systems deploying GPUs with significantly higher efficiency than CPU-based implementations, exemplified in our case study assessing the interaction energies in a benzene dimer. 
    \item {\it GPU support for MSD trial wavefunctions.} With customized CUDA kernels, we enabled an efficient realization of Wick's theorem.  Timing benchmarks for [Cu$_2$O$_2$]$^{2+}$ and [Fe$_2$S$_2$(SCH$_3$)$_4$]$^{2-}$ were shown to achieve more than an order of magnitude speedup compared to our CPU implementation for a large MSD trial.~\cite{msdipie}
    \item {\it Support for complex-valued Cholesky vectors.} \ipie{} can handle complex-valued Cholesky vectors that may arise when the underlying basis functions are complex-valued.
    \item {\it Free-projection AFQMC.} A numerically exact AFQMC approach can be used to study small strongly correlated systems.
    \item {\it Finite-temperature AFQMC.} A finite-temperature AFQMC algorithm based on the grand canonical ensemble was added.
    \item{\it Electron--phonon QMC.} A QMC algorithm that computes the ground state of electron--phonon problems was added.
    \item{\it Automatic differentiable AFQMC.} We offer AFQMC property calculations via automatic differentiation.
    \item{\it External package interfaces.} \ipie{} is now interfaced with PySCF, Dice, \trexio{}, and FQE. 
\end{enumerate}

\add{\texttt{ipie} has been mainly designed for ab initio calculations, 
    but more supports for model Hamiltonians have been added and are under active development,
    including the Hubbard-Holstein~\cite{lee2021constrained}, Peierls,
    and uniform electron gas models~\cite{lee2019auxiliary}, etc.
    The embedding method is also an interesting feature and currently not supported in \texttt{ipie}, 
    but it has been studied in literatures~\cite{PhysRevB.95.045103,Eskridge2019Jun},
    where AFQMC was employed as a solver for a model Hamiltonian
    in the embedding method.
    With the improved modularity and usability, we believe it would be relatively straightforward to implement.
    We also note that \texttt{ipie} currently uses double precision arithmetic for all calculations, although it also supports mixed precision where 
single precision is used for propagations and double precision is used for local energy estimation, 
the stability and efficiency need to be further investigated.
Also, there is indeed potential for further optimization in the CPU implementation, 
    particularly through enhancements with a hybrid MPI/OpenMP programming strategy,
    and we will continue to explore avenues along those lines.
} 

We hope that \ipie{} will serve as a community code base for developing {\it ab initio} AFQMC methods and their applications.
Furthermore, as \ipie{} is written mainly in Python, we anticipate its use in machine learning and quantum computing~\cite{Jiang2024Jul} communities will also grow.
\section{Acknowledgements}
The work by T.J. and J.L. was supported by the Department of Energy (DOE) Office of Fusion Energy Sciences “Foundations for quantum simulation of warm dense matter” project and by Harvard University's startup funds. Computations were carried out partly on the FASRC cluster supported by the FAS Division of Science Research Computing Group at Harvard University. This work also used the
Delta system at the National Center for Supercomputing Applications through allocation CHE230032, CHE230088, and PHY230192 from the
Advanced Cyberinfrastructure Coordination Ecosystem:
Services \& Support (ACCESS) program, which is supported by National Science Foundation grants \#2138259,
\#2138286, \#2138307, \#2137603, and \#2138296. 
P.F.L.~and A.S.~have received financial support from the European Research Council (ERC) under the European Union's Horizon 2020 research and innovation programme (Grant agreement no.~863481). A.S.~ was also supported by the European Centre of Excellence in Exascale Computing (TREX) which has received funding from the European Union's Horizon 2020 --- Research and Innovation program --- under grant agreement no.~952165.
S.F.U. thanks David Reichman for his support.
We thank Yifei Huang and Dingshun Lv for their GPU-MSD code contributions.~\cite{msdipie}

\section{Data Availability}\label{sec:data}
The data that support the findings of this study
are openly available in our Zenodo repository at
\hyperlink{https://doi.org/10.5281/zenodo.12522916}{https://doi.org/10.5281/zenodo.12522916}.

\bibliography{references}

\end{document}